\documentclass[pdflatex,sn-mathphys,Numbered]{sn-jnl}

\usepackage{graphicx}
\usepackage{subcaption}
\graphicspath{{figures/}}
\usepackage{letltxmacro}
\LetLtxMacro{\oldincludegraphics}{\includegraphics}
\renewcommand{\includegraphics}[2][]{%
  \IfFileExists{#2}{\oldincludegraphics[#1]{#2}}{%
    \IfFileExists{#2.pdf}{\oldincludegraphics[#1]{#2.pdf}}{%
      \IfFileExists{#2.eps}{\oldincludegraphics[#1]{#2.eps}}{%
        \fbox{Missing: \detokenize{#2}}%
      }%
    }%
  }%
}

\usepackage{amsmath}
\usepackage{amssymb}
\usepackage{amsthm}

\newcommand{\codename}{\textsc{Janus}}
\newcommand{\make}{\textsc{Make}}
\newcommand{\rmm}{\textsc{Rmm}}

\usepackage{booktabs}
\usepackage{tabularx}
\usepackage{multirow}
\usepackage{xcolor}
\usepackage{colortbl}
\usepackage{pifont}   

\usepackage{enumitem}
\setlist[itemize]{leftmargin=*, topsep=3pt, itemsep=1pt}
\setlist[enumerate]{leftmargin=*, topsep=3pt, itemsep=1pt}

\usepackage{balance}
\usepackage{manyfoot}

\usepackage{tikz}
\usetikzlibrary{arrows.meta, positioning, calc, shapes.geometric, patterns}
\usepackage{pgfplots}
\pgfplotsset{compat=1.16}

\usepackage{listings}
\def\make{\textsc{MAKE}}
\def\rmm{\textsc{RMM}}

\renewcommand{\checkmark}{\ding{51}}
\newcommand{\xmark}{\ding{55}}

\begin{document}

\title[Enforcing Attestable Workflows]{Enforcing Attestable Workflows across Untrusted Networks}

\author*[1]{\fnm{Hung} \sur{Dang}}\email{hung.dang@vlu.edu.vn}
\author[2]{\fnm{Tue} \sur{Nguyen}}\email{tuent@appota.com}

\affil*[1]{\orgdiv{Van Lang School of Technology}, \orgname{Van Lang University}, \orgaddress{\country{Vietnam}}}
\affil[2]{\orgname{Appota}, \orgaddress{\country{Vietnam}}}

\keywords{Confidential computing, Trusted execution environments, High-performance networking, eBPF, Intel TDX}

\abstract{Confidential high-performance computing orchestrates workloads across federated domains, yet existing frameworks rely on high-overhead user-space library operating systems or assume single-host execution. We propose \codename, an architecture federating Trusted Execution Environments via a split Trusted Computing Base (TCB) design. It couples a hardware-isolated Control Plane executing Mutually Attested Key Exchange (\make) with a measured guest-resident extended Berkeley Packet Filter (eBPF) Data Plane. By anchoring cryptographic key release to hardware measurements and executing enforcement in the kernel, \codename\ achieves native-speed encrypted routing. Empirical evaluation demonstrates a steady-state enforcement cost of $6\,\mu$s per packet, imposing a $13$--$15\,\mu$s absolute latency overhead. On distributed pipelines, \codename\ incurs just a $6.1\%$ execution penalty over plaintext baselines, bypassing the $62\%$ penalty of user-space counterparts. The system initializes a 100-node cluster in under 1.5 seconds, providing an efficient confidential interconnect for long-running workflows.}

\maketitle

\section{Introduction}
\label{sec:intro}

Modern high-performance computing (HPC) clusters and distributed computing platforms increasingly orchestrate confidential scientific workloads across federated administrative domains. Applications such as distributed medical imaging analysis, federated machine learning, and multi-site scientific simulations process highly sensitive datasets requiring strict privacy guarantees. Achieving multi-gigabit confidential networking under hardware-attested policies represents the central engineering challenge of confidential distributed workflows. While hardware primitives like Intel TDX~\cite{TDX} and Arm CCA~\cite{ARMCCA} provide strong local isolation for individual Confidential Virtual Machines (CVMs), they lack a distributed architecture to federate state and extend policy invariants across physical machine boundaries. In multi-site distributed contexts, data-flow scheduling and high-throughput communication patterns (e.g., pipeline DAGs~\cite{Bifrost2023}) must be cryptographically bound to platform attestation, ensuring intermediate data flows exclusively between verified enclave states.

Nonetheless, extending verifiable enforcement across untrusted network infrastructure faces various challenges. Standard cryptographic tunnels (e.g., TLS) secure data in transit but enforce no post-decryption behavioral policies, for an adversary controlling the destination host hypervisor can manipulate network namespaces or routing tables to redirect decrypted payloads to unauthorized enclaves without triggering network-level alarms. Plaintext gradient or data exchange across machine boundaries is therefore equivalent, from a privacy standpoint, to direct exposure. 

Confidential HPC networking must sustain 100GbE link rates with jumbo-frame MTUs while preserving end-to-end attestation invariants. Existing application-layer attestation frameworks (e.g., RA-TLS~\cite{RATLS}) embed cryptographic key release directly within the application computing base. While effective at establishing secure channels, this design imposes user-to-kernel boundary-crossing latency penalties and relies on the memory safety and protocol correctness of user-space libraries. Conversely, routing high-bandwidth socket streams through pure-firmware hypercalls incurs high context-switch overhead. Both approaches are incompatible with multi-gigabit data-plane throughput.

Whereas existing approaches perform enforcement either entirely in user-space or exclusively in firmware, we observe that the high-speed network routing path is independent from the complex, latency-tolerant remote attestation handshake. We propose \codename, an architecture constructing a Confidential Federated Interconnect that acts on this observation. Specifically, \codename\ utilizes a split-TCB architecture to balance hardware attestation with high-throughput routing. By anchoring enforcement in a measured guest-resident eBPF data plane, \codename\ transparently intercepts network traffic at the lowest layer of the kernel stack, while cryptographic keys are derived via an out-of-band hardware attestation protocol.

This split architecture provides three system-level advantages. First, it ensures completely transparent enforcement for unmodified UDP workloads, decoupling the cryptographic routing policy from the application logic entirely. Second, by expanding the Trusted Computing Base (TCB) to include the guest kernel, it shifts cryptographic enforcement from user-space libraries to a measured system-level boundary, enabling near-native throughput. Finally, distributed environments suffer from asynchronous policy drift during workflow updates. \codename\ prevents unauthorized communication during these divergent states through epoch-based policy versioning, rejecting handshakes lacking cryptographic consensus on the current policy digest.

We make the following contributions in this paper:
\begin{itemize}
\item \codename\ closes the \texttt{kTLS} attestation gap via \texttt{RTMR3}, BPF-Lock, and SEAM-gated key release, anchoring a split-TCB architecture that connects high-speed eBPF routing to hardware remote attestation.
\item Using Google Cloud Platform (GCP) Sapphire Rapids instances, empirical measurement of the static-key eBPF data plane establishes a bounded $6\,\mu$s per-packet processing cost. For stateless UDP workloads, isolated loopback microbenchmarks demonstrate this incurs a $13$--$15\,\mu$s absolute data-plane latency overhead over native unencrypted baselines, bypassing the LibOS syscall emulation overheads that drive the empirical $137$--$147\,\mu$s penalty of user-space RA-TLS.
\item We demonstrate macro-level scalability via a hybrid evaluation: empirical isolation of the data plane reveals a $6.1\%$ execution overhead over plaintext TCP on a distributed federated learning pipeline (ResNet-18), while natively measured attestation primitives analytically bound the integration cost (${+}{\sim}2.5\%$ median execution penalty per re-keying interval). By introducing lock-free epoch synchronization, we gracefully rotate cryptographic keys without disrupting in-flight UDP/FOU collective communications.
\item We implement the \make\ protocol to bind symmetric session keys to bilateral hardware states. Combining empirically measured per-handshake hardware costs with analytically grounded projections confirms a structured 100-node HPC pipeline initializes in under 1.5 seconds. We scope \codename's applicability to sparse topologies and long-running distributed pipelines, as dense $O(N^2)$ collectives expose serialization bottlenecks in current-generation hardware quoting enclaves.
\end{itemize}

The rest of this paper is organised as follows. Section~\ref{sec:background} provides background on Trusted Execution Environments and distributed confidential microservices. Section~\ref{sec:problem} formulates the problem and threat model. Section~\ref{sec:design} presents the \codename\ architecture and \make\ protocol. Section~\ref{sec:security} provides the security analysis. Section~\ref{sec:impl} describes the implementation. Section~\ref{sec:eval} presents the experimental evaluation, before Section~\ref{sec:conclusion} concludes the paper.

\section{Background and Related Work}
\label{sec:background}

\subsection{Trusted Execution Environments}
\label{sec:background:tee}

A Trusted Execution Environment (TEE) is a hardware-enforced isolated context protecting code and data confidentiality and integrity from privileged software (OS, hypervisor)~\cite{ARMCCA,TDX,SGX}. Early process-level TEEs like Intel SGX~\cite{SGX} isolate user-space processes. While offering a minimal Trusted Computing Base (TCB), they enforce strict memory limits (Enclave Page Cache) and often require specialized library operating systems to execute standard applications. Conversely, modern system-level TEEs like Intel TDX~\cite{TDX} and Arm CCA~\cite{ARMCCA} utilize memory encryption techniques, such as TDX's Multi-key Total-Memory Encryption (MKTME) coupled with the Secure Arbitration Mode (SEAM), to isolate entire Confidential Virtual Machines (CVMs) from the host hypervisor~\cite{FengSurvey}. This VM-level isolation allows CVMs to natively run unmodified guest operating systems, bypass library OS engineering overheads, and fully leverage the physical node's CPU and memory resources, drastically reducing the I/O and paging bottlenecks that plague large-memory workloads in SGX.

\textit{Remote attestation.}
TEEs produce cryptographically verifiable evidence of their state, allowing verifiers to confirm platform identity and enclave measurements~\cite{RATS,Sardar2024}. In Intel TDX~\cite{TDX}, the guest invokes a hypercall to generate a local \texttt{TDREPORT}, supplying a 64-byte \texttt{REPORTDATA} nonce to bind external data. The Intel-signed Quoting Enclave (QE) then transforms this into a remotely verifiable TD Quote.
\codename's \make\ protocol aligns with the IETF RATS architecture~\cite{RATS}, functioning as both Attester and Relying Party.  \make\ binds session keys directly to hardware-backed attestation reports. Sardar et al.~\cite{Sardar2024} identified initialization integrity violations in Arm CCA and Intel TDX using symbolic analysis, motivating our inclusion of the \make\ protocol directly in the measured \rmm\ image. ZKSA~\cite{ZKSA} targets TOCTOU attacks in IoT using zero-knowledge proofs, while PROVE~\cite{PROVE}, ZEKRA~\cite{ZEKRA}, and PRIVE~\cite{PRIVE} provide publicly verifiable, control-flow, and swarm attestation for external auditors. Conversely, \codename\ uses attestation primitives to directly gate network routing. 

\textit{RA-TLS.}
Integrating remote attestation with Transport Layer Security (RA-TLS)~\cite{RATLS} allows enclaves to embed attestation reports into TLS certificates, authenticating the peer and verifying hardware measurements simultaneously (pioneered by Knauth et al.~\cite{RATLS}). While \codename\ adopts this management channel abstraction, it moves enforcement from the application layer to the measured kernel-level data plane. Finally, device-level protocols like SPDM~\cite{SPDM} lack workflow-level policy enforcement. As noted by Feng et al.~\cite{FengSurvey}, distributed policy enforcement is a critical gap that \codename\ resolves in a platform-agnostic manner (independent of underlying OpenHCL~\cite{OpenHCL} or Graviton~\cite{Graviton} hardware advancements).

\subsection{Distributed Microservices in the Confidential Cloud}
\label{sec:background:distributed}

Cloud workloads distribute microservices across physical machines and administrative domains. While frameworks like Authentic Execution~\cite{Scopelliti2023} and RA-TLS~\cite{RATLS} extend TEE security to distributed settings, they require specific programming models or application-layer attestation. Systems isolating single CVMs (Erebor~\cite{Erebor}, Veil~\cite{Veil}) ignore cross-machine flows. Existing approaches place attestation and key management in the application layer, making confidentiality dependent on every workflow component's correctness. \codename\ elevates policy enforcement to a measured split-TCB level across boundaries, removing this dependency. 

\textit{Grid Frameworks and Intra-host Policies.}
Grid orchestration platforms like ParSL~\cite{babuji2019parsl} and HTCondor~\cite{thain2005distributed} abstract the complexity of executing massive Directed Acyclic Graphs (DAGs) across disparate resources. While they provide scheduling and fault tolerance, their security models assume a trusted underlying fabric or rely on centralized identity management, leaving data exposed in memory to compromised host operating systems. 

Conversely, recent Trusted Execution Environment advancements address intra-host data flow policies but struggle with distributed scale. The concurrent MICA~\cite{MICA} framework enforces local attestable data flow policies by decoupling confidentiality from application logic using physical memory isolation via the Reference Monitor Module. However, MICA solves an orthogonal problem domain: intra-host memory sharing among co-located enclaves on a single machine. It offers no remote verification mechanisms before data crosses machine boundaries. \codename\ specifically tackles the distributed challenge, securing inter-host data flows via cryptographic network bindings linking disparate physical hosts.

Other intra-host TEE systems include Veil~\cite{Veil} (nested VMMs for AMD SEV-SNP), Cerberus~\cite{Cerberus} (formalized SGX memory sharing), and Elasticlave~\cite{Elasticlave} (RISC-V shared memory). Sandboxing abstractions like InkTag~\cite{InkTag}, Erebor~\cite{Erebor}, and VeriSMo~\cite{VeriSMo} simplify or formalize intra-host CVM isolation. These works are complementary, strengthening single-host boundaries, whereas \codename\ establishes verifiable cross-machine enforcement suitable for wide-area grid orchestration.

\subsection{Distributed-Enforcement Approaches and Their Trade-offs}
\label{sec:background:approaches}

Several frameworks deploy TEEs across distributed settings. Ryoan~\cite{Ryoan} provides a distributed, data-oblivious sandbox. However, they require specific APIs or heavy application-layer instrumentation. \codename\ supports unmodified binaries by enforcing policy via a split architecture.
Kalium~\cite{Kalium} isolates serverless TEEs, and SRFL~\cite{Zhao2024CCPE} protects IoT federated learning models. HyperEnclave~\cite{HyperEnclave} abstracts TEE functionality across commodity hardware. We examine three closely related distributed-enforcement designs in detail, for each illuminates a distinct architectural axis along which \codename\ differs.

\emph{TEE Integration in HPC and Grid Environments.} Extending confidential computing into HPC contexts requires bridging hardware isolation with orchestration frameworks. Platforms like SCONE~\cite{SCONE} integrate SGX with container environments, providing a foundation for secure workflows but relying heavily on application-layer wrappers, while Veracruz~\cite{Veracruz} formalizes multi-party collaborative computations using a centralized policy broker.  \codename\ operates at the network stratum, directly mapping grid DAG topologies to decentralized hardware-attested endpoints without centralized coordination.

\emph{Authentic Execution}~\cite{Scopelliti2023} secures heterogeneous distributed enclave events through framework-specific event-binding logic embedded directly in the application runtime. The framework binds inter-component events to attestation evidence at the language-level boundary, requiring developers to structure applications around its event abstraction. \codename\ inverts this trade-off: it supports unmodified binaries by enforcing policy at a measured kernel-level boundary, accepting the TCB cost of including the guest kernel in exchange for application transparency. The two systems target complementary use cases: Authentic Execution suits new applications written against its event model, whereas \codename\ retrofits attested flow control onto unmodified UDP workloads.

\emph{Confidential Containers (CoCo) Attestation Agent} provisions decryption keys to container runtimes via a Key Broker Service after attesting the initial container state. Two architectural distinctions follow. First, CoCo attests the bootstrap state of the container, providing no point-to-point cryptographic binding between data flows and remote endpoint measurements; \codename's \make\ protocol binds each session key to bilateral hardware quotes per edge of the workflow DAG. Second, the synchronous KBS interaction imposes a 10--17\,s pod-startup tax that aggregates poorly across grid-scale workflows (though this tax is per-pod rather than per-flow and amortizes over long-running containers, it represents a rigid baseline for highly dynamic topologies); \codename\ amortizes \make\ across asynchronous parallel handshakes and projects sub-two-second initialization for the 100-node case (Section~\ref{sec:eval:scalability}).

\emph{In-kernel TLS (\texttt{kTLS})} with Linux's TLS upcall offload provides kernel-resident AES-GCM at near-native throughput but does not bind the symmetric key derivation to a measured network-stack state. Standard \texttt{kTLS} deployments derive keys via user-space libraries (e.g., OpenSSL, GnuTLS) and load them into the kernel through \texttt{setsockopt}, leaving the attestation evidence anchored to a user-space handshake while the encryption itself executes inside an unmeasured kernel. Even if the OpenSSL binary is measured into the RTMR during boot, the structural semantic gap remains: the key exchange terminates in user-space, and the kernel network stack receives the symmetric session keys blindly via the system call interface, possessing no cryptographic proof that those specific keys originated from an attested remote peer rather than a local user-space compromise. \codename's RTMR3 measurement of the eBPF/XDP proxy plus the BPF-Lock LSM closes this gap: key release is gated within the SEAM hardware boundary on the attested guest kernel state.

\section{Problem Formulation}
\label{sec:problem}

\subsection{Motivating Scenario}
\label{sec:problem:motivation}

Consider a distributed inference pipeline streaming sensitive queries (e.g., text for sentiment analysis) from ingest nodes to worker enclaves across distinct cloud providers. The workflow owner requires assurance that queries are processed only by a pre-approved BERT model enclave and forwarded exclusively to a designated aggregator enclave. If an intermediate host redirects data or an enclave is unauthorized, the workflow must fail without exposing data.

\begin{figure}[h]
  \centering
  \resizebox{\columnwidth}{!}{\begin{tikzpicture}[>=latex, font=\sffamily\small]
  \tikzstyle{box_host}=[draw=black, thick, fill=none, minimum width=2.8cm, minimum height=3.8cm, rounded corners]
  \tikzstyle{box_app}=[draw=blue!80!black, thick, fill=none, pattern color=blue!40, minimum width=2.2cm, minimum height=0.6cm]
  \tikzstyle{box_os_vuln}=[draw=red!80!black, thick, fill=none, pattern=dots, pattern color=red!40, minimum width=2.2cm, minimum height=0.6cm]
  \tikzstyle{box_os_safe}=[draw=gray!80!black, thick, fill=none, pattern=dots, pattern color=gray!60, minimum width=2.2cm, minimum height=0.6cm]
  \tikzstyle{box_rmm}=[draw=blue!60!black, thick, fill=none, pattern color=blue!60!black, minimum width=2.2cm, minimum height=0.6cm]

  \node at (0, 2.5) {\textbf{Application-Layer TLS}};
  
  \node[box_host] (hostA1) at (-1.8, 0) {};
  \node at (-1.8, 1.5) {Host A (Sender)};
  \node[box_app] (encA1) at (-1.8, 0.6) {App (TLS)};
  \node[box_os_safe] (osA1) at (-1.8, -0.8) {OS Kernel};

  \node[box_host] (hostB1) at (1.8, 0) {};
  \node at (1.8, 1.5) {Host B (Receiver)};
  \node[box_os_vuln] (encB1) at (1.8, 0.6) {App (TLS)};
  \node[box_os_vuln] (osB1) at (1.8, -0.8) {OS (Malicious)};

  \draw[->, thick, gray!60!black] (encA1) -- (osA1) node[midway, left=0.1cm, draw=gray!60!black, thick, rectangle, minimum width=0.4cm, minimum height=0.3cm, pattern=crosshatch, pattern color=gray!60!black] {};
  \draw[->, thick, gray!60!black] (osA1) -- (osB1) node[midway, above=0.1cm, draw=gray!60!black, thick, rectangle, minimum width=0.4cm, minimum height=0.3cm, pattern=crosshatch, pattern color=gray!60!black] {};
  \draw[->, thick, gray!60!black, transform canvas={xshift=-0.5cm}] (osB1) -- (encB1) node[midway, left=0.1cm, draw=gray!60!black, thick, rectangle, minimum width=0.4cm, minimum height=0.3cm, pattern=crosshatch, pattern color=gray!60!black] {};
  
  \draw[->, thick, red!60!black, transform canvas={xshift=0.5cm}] (encB1) -- (osB1)  node[midway, right=0.1cm, draw=gray!60!black, thick, rectangle, minimum width=0.4cm, minimum height=0.3cm, pattern color=gray!60!black] {};

  \draw[thick, dotted, gray] (4.0, -2.5) -- (4.0, 3); 

  \node at (8, 2.5) {\textbf{\codename\ Split-TCB}};
  
  \node[box_host] (hostA2) at (6.2, 0) {};
  \node at (6.2, 2.15) {Host A};
  \node[box_app] (encA2) at (6.2, 1.25) {Enclave App};
  \node[box_rmm] (rmmA2) at (6.2, 0.0) {\rmm\ (XDP)};
  \node[box_os_safe] (osA2) at (6.2, -1.2) {Untrusted OS};

  \node[box_host] (hostB2) at (9.8, 0) {};
  \node at (9.8, 2.15) {Host B};
  \node[box_app] (encB2) at (9.8, 1.25) {Enclave App};
  \node[box_rmm] (rmmB2) at (9.8, 0.0) {\rmm\ (XDP)};
  \node[box_os_vuln] (osB2) at (9.8, -1.2) {Untrusted OS};

  \draw[->, thick, blue!80!black] (encA2) -- (rmmA2) node[midway, left=0.1cm, draw=blue!80!black, thick, rectangle, minimum width=0.4cm, minimum height=0.3cm, fill=white] {};
  \draw[->, thick, gray!60!black] (rmmA2) -- (osA2) node[midway, left=0.1cm, draw=gray!60!black, thick, rectangle, minimum width=0.4cm, minimum height=0.3cm, pattern=crosshatch, pattern color=gray!60!black] {};
  \draw[->, thick, gray!60!black] (osA2) -- (osB2) node[midway, above=0.1cm, draw=gray!60!black, thick, rectangle, minimum width=0.4cm, minimum height=0.3cm, pattern=crosshatch, pattern color=gray!60!black] {};
  \draw[->, thick, gray!60!black] (osB2) -- (rmmB2) node[midway, right=0.1cm, draw=gray!60!black, thick, rectangle, minimum width=0.4cm, minimum height=0.3cm, pattern=crosshatch, pattern color=gray!60!black] {};
  
    \draw[->, thick, blue!80!black] (rmmB2) -- (encB2) node[midway, right=0.1cm, draw=blue!80!black, thick, rectangle, minimum width=0.4cm, minimum height=0.3cm, fill=white] {};
  
\end{tikzpicture}}
  \caption{The Distributed Security Gap. Left: Application-layer TLS of Host B was exploited. The exploited App leaks decrypted data to the malicious OS. Right: \codename's Split-TCB architecture intercepts traffic at the RMM boundary, strictly tying decryption to hardware-attested policy states before data reaches the enclave. Crosshatched rectangles depict encrypted data, whereas unfilled rectangles indicate plaintext.}
  \label{fig:security_gap}
\end{figure}
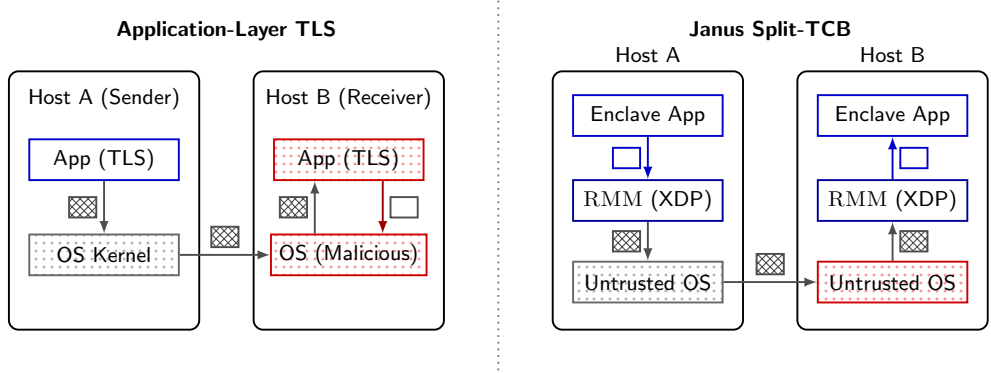

Figure~\ref{fig:security_gap} illustrates the fundamental vulnerability. While hardware execution environments protect memory from direct OS inspection, traditional application-layer TLS forces the application to handle decryption. This integrates complex software frameworks into the Trusted Computing Base (TCB). A malicious host OS can exploit vulnerabilities within this bloated application layer via crafted network payloads. Once compromised, the application decrypts the payload and leaks it back to the OS, bypassing hardware protections. A split-TCB architecture mitigates this by intercepting traffic at the \rmm\ boundary, strictly enforcing attested decryption before data reaches the vulnerable application space.

\subsection{System Model}
\label{sec:problem:system}

We model a distributed workflow as a directed acyclic graph $\mathcal{G} = (V, E)$ over compute nodes $v \in V$. Each edge $(v_i, v_j) \in E$ is a permitted data flow. Nodes execute on distinct, potentially cross-administrative physical hosts $h(v)$.

\textit{Client.}
The data owner defines $\mathcal{G}$ and the data flow policy $\mathcal{P}$, specifying expected destination enclave measurements and network address bindings for each edge. The client distributes the signed policy to all Reference Monitor Modules before execution.

\textit{Host.}
The host OS provides computing resources but is explicitly excluded from the Trusted Computing Base (TCB).

\textit{Reference Monitor Module (\rmm) and Data Plane.}
The \rmm\ is a privileged firmware component enforcing local policies. In Intel TDX, the \rmm\ extends the SEAM-mode TDX module, securely bootstrapping a guest-resident data plane. \codename\ expands the TCB as motivated in Section~\ref{sec:intro}. To ensure this TCB expansion does not expose the system to a malicious hypervisor supplying a compromised kernel, \codename\ relies on a strict secure boot chain: the Trust Domain Virtual Firmware (TDVF) cryptographically measures the Guest OS kernel, initial RAM disk (initrd), and boot parameters into the Run-Time Measurement Registers (\textit{RTMR1} and \textit{RTMR2}) before execution. This provides an execution path cryptographically bound to hardware attestation.

\textit{Trusted Execution Environment.}
Each node $v$ executes unmodified inside a TEE. Its initial memory state is captured in a cryptographic measurement $\mu(v)$ within the hardware attestation report. Network communication occurs over an asynchronous, untrusted wide-area network.

\subsection{Threat Model}
\label{sec:problem:threat}

We consider a PPT adversary $\mathcal{A}$ controlling the network infrastructure and host OSs. $\mathcal{A}$ can intercept, modify, replay, or drop packets; manipulate the hypervisor and redirect sockets via root privileges; deploy unauthorized enclaves; and observe all network traffic.
Traditional network-level controls fail against such host-level compromise. However, we trust the hardware processors, the manufacturer's attestation infrastructure, and the TDX module post-initialization~\cite{Sardar2024}. Hardware cryptographic keys are assumed unforgeable~\cite{RATS,Sardar2024}. 

We leave availability attacks (e.g., network-level denial-of-service or packet dropping) and traffic analysis (e.g., inferring workflow topologies via volume or timing side-channels) out of scope. Mitigating these threats requires orthogonal network-layer padding and Byzantine-fault-tolerant routing mechanisms. Side-channel attacks against the CPU microarchitecture (e.g., transient execution) remain outside our scope. \codename\ protects against network-level adversaries and untrusted hypervisors, but offers no defense against post-boot kernel-resident exploitation by a malicious application tenant.

\subsection{Security Goals}
\label{sec:problem:goals}

\codename\ must achieve the following properties:

\begin{itemize}
    \item \textbf{Policy Adherence:} Data $d$ governed by $\mathcal{P}$ must only flow along authorized edges $(v_i, v_j) \in E$. Destination node $v_j$ must present an attestation measurement $\mu(v_j)$ matching $\mathcal{P}$.
    \item \textbf{Data Confidentiality:} Data $d$ must never be accessible in plaintext to the network, untrusted host OS, or unauthorized enclaves during transit. Plaintext exposure is confined to the origin and destination boundaries, including their respective trusted (but vulnerable) guest OS instances. We note that this property governs network transit; mitigating post-decryption exfiltration via non-network covert channels (e.g., shared memory, cache timing) within the destination host remains outside the scope of this threat model.
    \item \textbf{Attested Flow Control:} The enforcement mechanism must cryptographically bind transmission channels to verifiable hardware states. A session key for $(v_i, v_j)$ must be bound to both attested measurements $\mu(v_i)$ and $\mu(v_j)$.
\end{itemize}

These properties are compositional: if Policy Adherence, Data Confidentiality, and Attested Flow Control hold for every edge and transit segment, the workflow provides end-to-end confidentiality.

\section{System Design: \codename}
\label{sec:design}

\subsection{High-Level Overview}
\label{sec:design:overview}

\begin{figure}[t]
  \centering
  \resizebox{\columnwidth}{!}{\begin{tikzpicture}[>=latex, x=2cm, y=1.2cm, font=\sffamily\small]

  \draw[fill=none, draw=none] (-0.5, 4.5) rectangle (6.5, 5.5);
  \node[left, text=blue!80!black] at (-0.6, 5.0) {Guest Userspace};
  
  \draw[fill=none, draw=none] (-0.5, 3.5) rectangle (6.5, 4.5);
  \node[left, text=green!60!black] at (-0.6, 4.0) {Guest Kernel};
  
  \draw[fill=none, draw=none] (-0.5, 2.5) rectangle (6.5, 3.5);
  \node[left, text=purple!80!black] at (-0.6, 3.0) {SEAM Mode};
  
  \draw[fill=none, draw=none] (-0.5, 1.5) rectangle (6.5, 2.5);
  \node[left, text=red!80!black] at (-0.6, 2.0) {Untrusted Host};

  \draw[fill=none, draw=none] (-0.5, 0.5) rectangle (6.5, 1.5);
  \node[left, text=gray!80!black] at (-0.6, 1.0) {Network};

  \foreach \y in {1.5, 2.5, 3.5, 4.5} {
    \draw[dash dot dot, gray] (-0.5, \y) -- (6.5, \y);
  }

  \node[draw=blue!80!black, thick, fill=none, minimum width=2.5cm, minimum height=0.6cm, pattern color=blue!40] (appA) at (1, 5.0) {Application (Ring-3)};
  
  \node[draw=green!60!black, thick, fill=none, minimum width=2.5cm, minimum height=0.6cm, pattern color=green!60] (dpA) at (1, 4.0) {eBPF Data Plane};
  
  \node[draw=purple!80!black, thick, fill=none, minimum width=2.5cm, minimum height=0.6cm, pattern color=purple!40] (rmmA) at (1, 3.0) {\rmm\ Control Plane};
  
  \node[draw=red!80!black, thick, fill=none, minimum width=2.5cm, minimum height=0.6cm, pattern=dots, pattern color=red!40] (osA) at (1, 2.0) {Host OS};

  \node[draw=blue!80!black, thick, fill=none, minimum width=2.5cm, minimum height=0.6cm,  pattern color=blue!40] (appB) at (5, 5.0) {Application (Ring-3)};
  
  \node[draw=green!60!black, thick, fill=none, minimum width=2.5cm, minimum height=0.6cm,  pattern color=green!60] (dpB) at (5, 4.0) {eBPF Data Plane};
  
  \node[draw=purple!80!black, thick, fill=none, minimum width=2.5cm, minimum height=0.6cm,  pattern color=purple!40] (rmmB) at (5, 3.0) {\rmm\ Control Plane};
  
  \node[draw=red!80!black, thick, fill=none, minimum width=2.5cm, minimum height=0.6cm, pattern=dots, pattern color=red!40] (osB) at (5, 2.0) {Host OS};

  \draw[->, thick, blue!80!black] (appA) -- (dpA) node[midway, right, font=\small] {};
  \draw[->, thick, blue!80!black] (dpB) -- (appB) node[midway, left, font=\tiny] {};
  
  \draw[<->, thick, orange!80!black] (dpA) -- (rmmA) node[midway, right, font=\small, align=left] {};
  \draw[<->, thick, orange!80!black] (dpB) -- (rmmB) node[midway, left, font=\tiny, align=right] {};
  
  \draw[<->, thick, purple!80!black] (rmmA.east) -- (rmmB.west) node[midway, above = -0.05cm, font=\small] {\make\ Handshake};
  
  
  \draw[->, thick, green!60!black, dashed] (dpA.south) ++ (-0.23,0) -- ++(0,-1.6) coordinate (pA) -- (pA -| osB.west) node[pos = 0.6, above=-0.05cm, font=\small] { Encrypted Bulk Traffic} ++ (0.386,0) -- ++(0,1.7);

  \node[draw=black, fill=white, thick, minimum width=2cm, minimum height=0.6cm, rounded corners] (net) at (3.0, 1.0) {Internet/WAN};
  
  \draw[<->, thick, gray] (osA) -- (osA |- net.north);
  \draw[<->, thick, gray] (osB) -- (osB |- net.north);

\end{tikzpicture}}
  \caption{\codename\ architecture. The \rmm\ operates as a SEAM-mode Control Plane negotiating an attested session with the remote \rmm. The measured Guest-Resident Data Plane transparently enforces encrypted flows before network release.}
  \label{fig:architecture}
\end{figure}
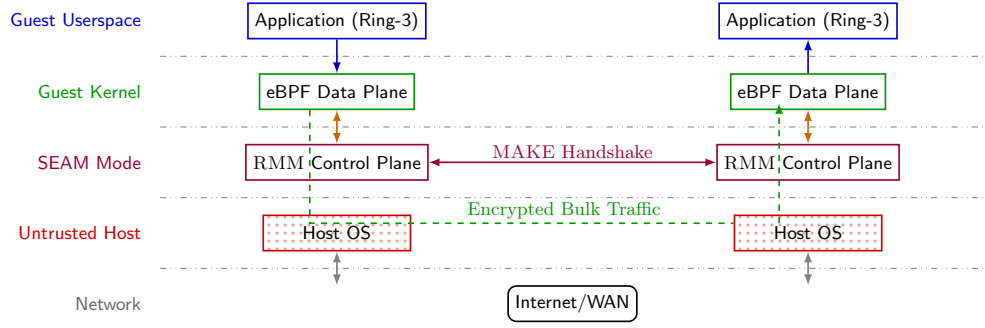

\codename\ transforms isolated Reference Monitor Modules (\rmm s) into a unified distributed enforcement mechanism (Figure~\ref{fig:architecture}). Clients sign and distribute workflow policies $\mathcal{P}$ to all participating \rmm s, which validate the signatures and install flow rules. 
Instead of trusting the host network stack, \codename\ intercepts outbound network operations via a split architecture. When connecting to an authorized endpoint, the source \rmm\ initiates a Mutually Attested Key Exchange (\make) session with the destination \rmm. This yields a symmetric session key conditioned on bilateral hardware attestations. Subsequent socket data is transparently encrypted/decrypted by the \rmm\ data planes, blinding the untrusted host.
\codename\ requires no application modifications for native UDP flows. To support TCP-bound orchestration frameworks (see Section~\ref{sec:eval:macro}), \codename\ requires operator-configured stateless Linux Foo-over-UDP (FOU) encapsulation, representing a deliberate deployment mode that preserves end-to-end TCP semantics over UDP datagrams but necessitates MTU tuning. As detailed in Section~\ref{sec:intro}, \codename\ avoids the prohibitive latency of pure-firmware enforcement via a hybrid split architecture: a SEAM-mode Control Plane (\rmm) handles \make, key provisioning, and policy attestation, while a securely bootstrapped, measured Guest-Resident eBPF Data Plane transparently intercepts socket operations.

\subsection{Split Control/Data Plane and Trust Establishment}
\label{sec:design:split}
\codename's TCB encompasses the SEAM-mode \rmm\ (Control Plane), the Guest OS Kernel, and the eBPF/XDP proxy (Data Plane). The host OS is excluded.
To prevent Time-Of-Check to Time-Of-Use (TOCTOU) attacks, the Data Plane is cryptographically bound to the TDX hardware state. During initialization, the Guest OS Kernel extends the eBPF/XDP proxy measurement into the TDX Run-Time Measurement Register 3 (\texttt{RTMR3}).
\begin{figure}[t]
  \centering
  \resizebox{\columnwidth}{!}{\begin{tikzpicture}[x=2cm, y=1cm, font=\sffamily\small]

  \draw[->, thick] (0, 0) -- (7, 0) node[right] {Boot Time};
  
  \foreach \x in {0.5, 2.0, 3.5, 5.0, 6.5} {
    \draw (\x, -0.2) -- (\x, 0.2);
  }

  \node[above, align=center, yshift=0.25cm] at (0.5, 0) {\textbf{systemd-init} \\ Execution};
  \node[above, align=center, yshift=0.25cm] at (2.0, 0) {Load XDP \\ Proxy};
  \node[above, align=center, yshift=0.25cm] at (3.5, 0) {Activate \\ BPF-Lock LSM};
  \node[above, align=center, yshift=0.25cm] at (5.0, 0) {Extend \\ \texttt{RTMR3}};
  \node[above, align=center, yshift=0.25cm] at (6.5, 0) {Pivot to \\ Root FS};

  \draw[<->, draw=red!80!black, thick] (2.0, -0.5) -- (3.5, -0.5) node[midway, below, text=red!80!black] {\textbf{TOCTOU Window}};
  \draw[dashed, red!80!black] (2.0, 0) -- (2.0, -0.45);
  \draw[dashed, red!80!black] (3.5, 0) -- (3.5, -0.45);

  \draw[decorate,decoration={brace,amplitude=5pt,mirror},thick,draw=blue!80!black] 
    (0.4, -1) -- (5.1, -1) node[midway, below=6pt, text=blue!80!black] {Measured \texttt{initrd} (No user-space)};
    
  \draw[decorate,decoration={brace,amplitude=5pt,mirror},thick,draw=green!60!black] 
    (5.2, -1) -- (6.8, -1) node[midway, below=6pt, text=green!60!black] {Untrusted User-space};

\end{tikzpicture}}
  \caption{Early-boot sequence establishing the measured environment. The TOCTOU window is strictly closed prior to pivoting to the root filesystem, precluding unprivileged user-space interference.}
  \label{fig:boot_sequence}
\end{figure}
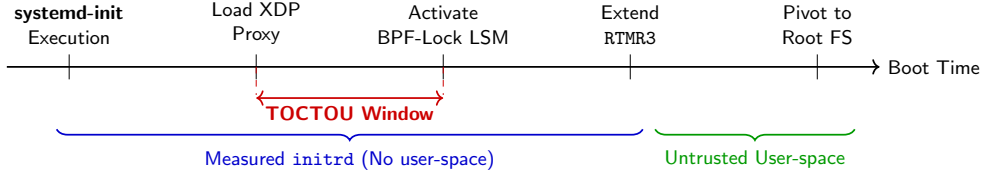

When the Data Plane requests session keys via a \texttt{TDCALL}, the \rmm\ verifies the guest's \texttt{RTMR3} against the expected measurement in $\mathcal{P}$. Keys and routing rules are provisioned only on a match. Providing defense-in-depth against post-boot guest-kernel exploitation, \codename\ requires the Guest OS to use a custom BPF Linux Security Module (LSM) program that globally disables the \texttt{bpf(BPF\_PROG\_LOAD)} syscall after loading the authorized XDP proxy. This ``BPF-Lock'' state is also extended into \texttt{RTMR3}. This prevents the loading of new BPF programs after boot, closing one TOCTOU vector. We note that the expected \texttt{RTMR3} sequence is computed dynamically prior to deployment via a trusted CI/CD pipeline, intrinsically binding it to exact kernel and LSM versions. As illustrated in Figure~\ref{fig:boot_sequence}, the small TOCTOU window between proxy load and BPF-Lock activation necessitates atomic execution during early-boot to prevent rogue program injection. Specifically, the measured \texttt{initrd} executes an early-boot \texttt{systemd} unit that loads the eBPF proxy and immediately activates the LSM lock before pivoting to the root filesystem; this guarantees no unprivileged user-space processes exist to invoke \texttt{bpf()} during this window. However, comprehensive runtime kernel integrity against direct memory corruption or ROP payloads remains an open challenge inherited from the Linux kernel hardening state of the art. While an attacker with root privileges can trigger a panic to cause a denial-of-service, \codename\ prioritizes safety over liveness; any detected policy violation or TCB compromise triggers a kernel panic, effectively halting the data flow to prevent leakage at the cost of service availability.

\subsection{Component A: Extended Policy Language and Versioning}
\label{sec:design:policy}

\codename\ utilizes a declarative JSON policy model featuring an \texttt{Epoch} version counter and a \texttt{RemotePeers} section. Each remote peer entry specifies the expected IP address, the SHA-384 measurement $\mu(v_j)$, and the expected \texttt{policy\_digest}~$\pi$ (SHA-384 hash of the policy rules and \texttt{Epoch} ID).
To handle distributed policy drift, each policy bundle includes a monotonically increasing \texttt{Epoch} ID. Handshakes proceed only if both endpoints enforce the same epoch; mismatches trigger asynchronous client updates, ensuring synchronized enforcement.

\subsection{Component B: Mutually Attested Key Exchange}
\label{sec:design:make}

\begin{figure}[t]
  \centering
  \resizebox{0.85\columnwidth}{!}{\begin{tikzpicture}[>=latex, font=\sffamily\fontsize{6pt}{7pt}\selectfont, scale=0.75, transform shape]
  \node (rs) at (0, 0) [draw=blue!80!black, thick, fill=none, pattern color=blue!20, minimum width=2.2cm, minimum height=0.6cm, rounded corners] {Source $\text{RMM}_s$};
  \node (rd) at (8, 0) [draw=blue!80!black, thick, fill=none,pattern color=blue!20, minimum width=2.2cm, minimum height=0.6cm, rounded corners] {Dest $\text{RMM}_d$};
  \draw[thick, dashed] (rs.south) -- (0, -5.4);
  \draw[thick, dashed] (rd.south) -- (8, -5.4);

  \node[anchor=west, align=left, font=\fontsize{6pt}{7pt}\selectfont] at (0.2, -0.55) {1. Gen $(sk_s, pk_s)$, nonce $n_s$};
  \node[anchor=west, align=left, font=\fontsize{6pt}{7pt}\selectfont] at (0.2, -0.95) {2. \texttt{ReportData} $\leftarrow H(pk_s, \pi_s, n_s)$};
  \node[anchor=west, align=left, font=\fontsize{6pt}{7pt}\selectfont] at (0.2, -1.35) {3. Get TD Quote $Q_s$};

  \draw[->, thick, blue!80!black] (0, -1.8) -- (8, -1.8) node[midway, above, font=\fontsize{6pt}{7pt}\selectfont] {4. Flight 1: $(pk_s, Q_s, n_s)$};

  \node[anchor=east, align=left, font=\fontsize{6pt}{7pt}\selectfont] at (7.8, -2.25) {5. Verify $Q_s$, check $\mu_s, \pi_s \in \mathcal{P}$};
  \node[anchor=east, align=left, font=\fontsize{6pt}{7pt}\selectfont] at (7.8, -2.65) {6. Gen $(sk_d, pk_d)$, nonce $n_d$};
  \node[anchor=east, align=left, font=\fontsize{6pt}{7pt}\selectfont] at (7.8, -3.05) {\ \ \texttt{ReportData} $\leftarrow H(pk_d, \pi_d, n_d, \mathbf{pk_s}, \mathbf{n_s})$};
  \node[anchor=east, align=left, font=\fontsize{6pt}{7pt}\selectfont] at (7.8, -3.45) {\ \ Get TD Quote $Q_d$};

  \draw[<-, thick, blue!80!black] (0, -3.9) -- (8, -3.9) node[midway, above, font=\fontsize{6pt}{7pt}\selectfont] {7. Flight 2: $(pk_d, \mathbf{\pi_d}, Q_d, n_d)$};

  \node[anchor=west, align=left, font=\fontsize{6pt}{7pt}\selectfont] at (0.2, -4.35) {8. Verify $Q_d$, confirm $\mathbf{pk_s}, \mathbf{n_s}$ reflected};

  \draw[->, thick, blue!80!black] (0, -4.85) -- (8, -4.85) node[midway, above, font=\fontsize{6pt}{7pt}\selectfont] {9. Flight 3: $\text{MAC}_K(Q_d \parallel Q_s)$};

  \node[anchor=east, align=right, font=\fontsize{6pt}{7pt}\selectfont] at (7.8, -5.3) {10. Verify $\text{MAC}_K$};

  \node[draw=green!60!black, thick, fill=none, rounded corners, minimum width=8.5cm, align=center, pattern color=green!20, font=\fontsize{6pt}{7pt}\selectfont, inner sep=3pt] (algebra) at (4, -6.3) {
    \textbf{Shared Session Key Derivation (HKDF-ECDH)} \\
    $R_s$: $K \leftarrow \text{HKDF}(\text{ECDH}(sk_s, pk_d), Q_s \parallel Q_d)$ \\
    $R_d$: $K \leftarrow \text{HKDF}(\text{ECDH}(sk_d, pk_s), Q_s \parallel Q_d)$
  };
  \draw[thick, dashed, green!60!black] (0, -5.4) -- (0, -5.85) -- (algebra.north -| 0, -5.85);
  \draw[thick, dashed, green!60!black] (8, -5.4) -- (8, -5.85) -- (algebra.north -| 8, -5.85);

\end{tikzpicture}}
  \caption{The \make\ protocol sequence. $R_d$'s \texttt{ReportData} (Step 6) securely binds the handshake to $pk_s$.}
  \label{fig:make-protocol}
\end{figure}
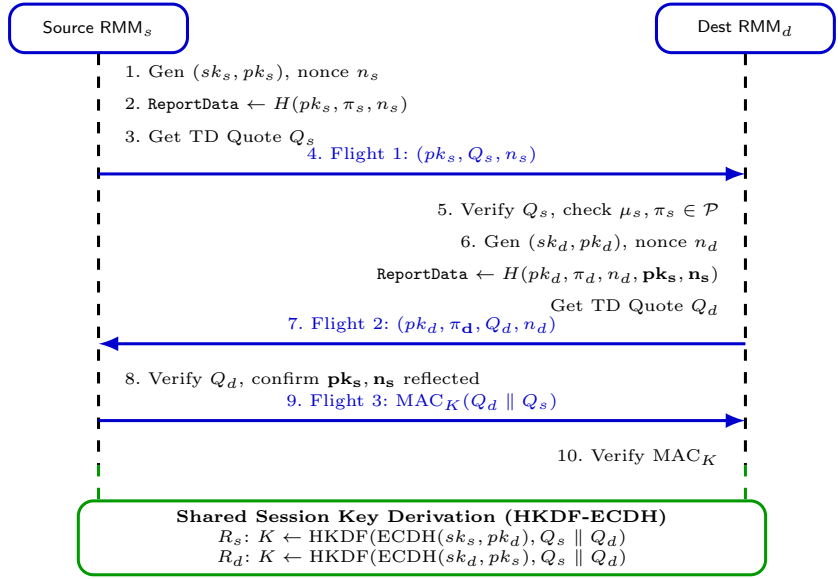

The \make\ protocol establishes a symmetric session key between \rmm s $R_s$ and $R_d$, conditioned on valid mutual attestation (Figure~\ref{fig:make-protocol}).

\textit{Cryptographic Binding and the SIGMA-I Pattern.}
\make\ follows the SIGMA-I pattern (Sign-and-MAC, Initiator privacy)~\cite{Krawczyk2003}. The responder ($R_d$) binds its attestation to the initiator's parameters by including the initiator's public key $pk_s$ in its attested data (Step 6), preventing man-in-the-middle attacks. This requirement forces the responder to generate a fresh TD Quote per handshake. Since the TDX \texttt{ReportData} requires 64 bytes, the binding right-pads the big-endian SHA-384 digest of handshake parameters ($pk_d, \pi_d, n_d, pk_s, n_s$) with 16 zero bytes. The initiator ($R_s$) symmetrically binds the responder's public key $pk_d$ by transmitting an authenticated MAC back to $R_d$ in a final third flight (Step 9), ensuring cryptographically bound sessions for both endpoints. Fresh 256-bit hardware-sampled nonces ($n_s, n_d$) prevent replay attacks.

\subsection{Component C: Distributed Policy Consistency}
\label{sec:design:consistency}

\begin{figure}[t]
  \centering
  \begin{subfigure}[b]{0.48\columnwidth}
    \centering
    \resizebox{\linewidth}{!}{\begin{tikzpicture}[->, >=stealth, auto, node distance=3.5cm, semithick, font=\sffamily\small]
  \tikzstyle{state}=[rectangle, rounded corners, draw=black, thick, text width=2.5cm, align=center, minimum height=1.2cm, fill=none]

  \node[state] (active) {Epoch $E$ \\ \textbf{Active}};
  \node[state] (provisioned) [right of=active, node distance=5cm] {Epoch $E+1$ \\ \textbf{Provisioned}};
  \node[state] (grace) [below of=provisioned] {Epoch $E$ \\ \textbf{Grace Period}};
  \node[state] (flushed) [left of=grace, node distance=5cm] {Epoch $E$ \\ \textbf{Stale-Flushed}};

  \path (active) edge node[above, align=center] {Policy $E+1$ \\ provisioned} (provisioned)
        (provisioned) edge node[right, align=left] {Mark $E$ stale \\ \make\ requires $E+1$} (grace)
        (grace) edge node[below, align=center] {Queue drains \\ RCU expires} (flushed)
        (flushed) edge[dashed, bend left=20] node[left, align=center] {Keys destroyed \\ ($E \leftarrow E+1$)} (active);
        

\end{tikzpicture}}
    \caption{Per-node key lifecycle. Cryptographic invariants bind each edge, ensuring seamless rotation without data leaks.}
    \label{fig:epoch_state_machine}
  \end{subfigure}
  \hfill
  \begin{subfigure}[b]{0.48\columnwidth}
    \centering
    \resizebox{\linewidth}{!}{\begin{tikzpicture}[>=latex, font=\sffamily\scriptsize, scale=0.7, transform shape]
  \node (client) at (0, 0) [draw=orange!80!black, thick, fill=none, pattern color=orange!20, minimum width=1.6cm, minimum height=2cm, rounded corners, align=center] {Client\\(Policy\\Owner)};

  \node (rmmA) at (5, 1.0) [draw=purple!80!black, thick, fill=none, pattern color=purple!20, minimum width=2.1cm, minimum height=0.8cm, rounded corners, align=center] {$\text{RMM}_A$\\ \fontsize{6pt}{7pt}\selectfont Epoch $N+1$};
  \node (rmmB) at (5, -1.0) [draw=purple!80!black, thick, fill=none,  pattern color=purple!20, minimum width=2.1cm, minimum height=0.8cm, rounded corners, align=center] {$\text{RMM}_B$\\ \fontsize{6pt}{7pt}\selectfont Epoch $N$};

  \draw[->, thick, dashed, blue!80!black] (client.east) -- (rmmA.west) node[midway, above, font=\fontsize{6pt}{7pt}\selectfont, align=center, sloped] {1. Push Policy $V_{N+1}$};
  \draw[->, thick, dashed, gray] (client.east) -- (rmmB.west) node[midway, below, font=\fontsize{6pt}{7pt}\selectfont, align=center, sloped] {(Delayed update)};

  \draw[->, thick, red!80!black] (rmmA.south) -- (rmmB.north) node[midway, right, font=\fontsize{6pt}{7pt}\selectfont, align=left] {2. \make\ Handshake\\(Offers Epoch $N+1$)};


 \node (err) at (5.8, -2.1) [text=brown!80!black, align=center] {Epoch \\Mismatch};
 \draw[->, thick, blue!80!black] 
    (rmmB.south) |- 
    ([yshift=-1cm]rmmB.south) -| 
    node[pos=0.25, below, font=\fontsize{6pt}{7pt}\selectfont, align=center] {3. Async Sync\\Request $V_{N+1}$} 
    (client.south);

\end{tikzpicture}}
    \caption{Cross-node epoch reconciliation. \make\ prevents communication under divergent policy states.}
    \label{fig:epoch_sync}
  \end{subfigure}
  \caption{Distributed policy consistency. (a) The lock-free state machine governs intra-node key rotation across an RCU grace period; (b) the cross-node \make\ handshake detects epoch mismatches and triggers asynchronous client-driven synchronization.}
  \label{fig:epoch_consistency}
\end{figure}
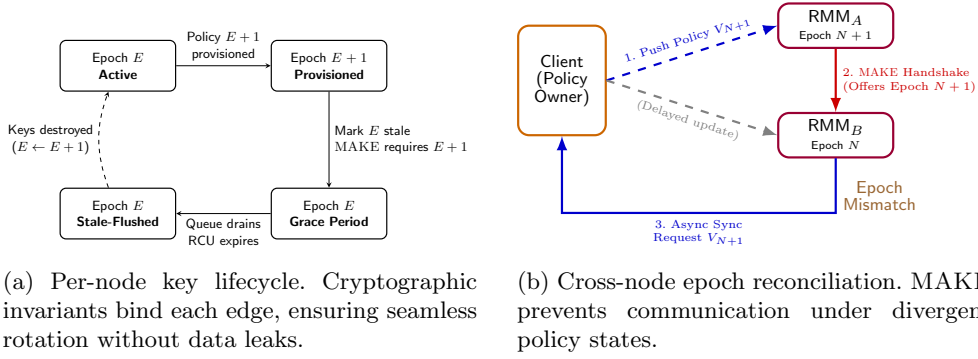

\codename\ employs a two-phase invalidation strategy to enforce strict distributed policy consistency across untrusted boundaries (Figure~\ref{fig:epoch_consistency}). When a workflow provisions version $N+1$, the \rmm\ immediately marks version $N$ keys as stale. To facilitate graceful key rotation without disrupting in-flight multi-gigabyte MPI transfers or forcing TCP connection resets, epoch rollover updates a lock-free per-CPU map. This mechanism deterministically isolates race conditions by tracking explicit RCU (Read-Copy-Update) grace periods; bytes already buffered in the network stack process under the stale key until the queue drains and the grace period expires, after which the eBPF filter invalidates the session and zeroes the stale keys from memory. 
New connections strictly mandate version $N+1$ \make\ handshakes. Epoch mismatches immediately trigger synchronization (Step 5 of Figure~\ref{fig:make-protocol}), preventing nodes from executing divergent security assumptions. An adversary cannot replay an Epoch $N-1$ handshake to a node running Epoch $N+1$; the expected policy digest $\pi$ embedded in the hardware quote differs, and the inherent monotonicity of the epoch counter rejects policy downgrades.

\textit{Node Churn and Fault Tolerance.}
Because \codename\ operates statelessly at the data plane, node churn and pipeline failures are handled automatically. The system inherits the fail-stop semantics of TDX; compromised or failed nodes disconnect deterministically, and mid-workflow crashes result in transport-layer timeouts. Recovery requires instantiating a replacement CVM and completing a fresh \make\ handshake ($103.2$\,ms). This mechanism imposes negligible recovery latency relative to typical scientific pipeline checkpoint-restart overheads. Replacing a physical TDX node generates a new cryptographic measurement, requiring the client to explicitly authorize recovery by distributing a new policy epoch.

\subsection{Hardware-Agnostic Architecture}
\label{sec:design:hardware}

While our prototype implementation strictly targets Intel TDX utilizing \texttt{TDCALL} and \texttt{SEAMCALL} interfaces, the split-TCB enforcement and attestation-bound key exchange principles remain hardware-agnostic. Porting \codename\ to Arm CCA requires mapping the Control Plane to the EL2 Realm Management Monitor (RMM), utilizing Realm Extensible Measurements (REM) instead of \texttt{RTMR3}, and binding to the Realm Service Interface (RSI) via Entity Attestation Tokens (EAT). Similarly, extending the architecture to AMD SEV-SNP entails mapping the Control Plane to the Secure VM Service Module (SVSM) operating at VMPL0, utilizing the \texttt{MEASURE} interface for guest extensions, and invoking \texttt{SNP\_GUEST\_REQUEST} for hardware attestation. Published AMD SEV-SNP attestation latencies ($\sim$2--6\,ms~\cite{SEVSNP_Perf}) suggest cross-platform deployments would eliminate the Intel Quoting Enclave serialization bottleneck entirely, demonstrating the broad applicability of the split-TCB design.

\section{Security Analysis}
\label{sec:security}

We analyze \codename's architectural resilience against the infrastructure-level adversaries defined in Section~\ref{sec:problem:threat}, systematically decomposing the systems mechanisms that satisfy our specified security goals.

\begin{table}[h]
\centering
\caption{Guest OS hardening mitigations mapping to mitigated attack classes. This explicit configuration dictates the security baseline required to validate the measured Guest OS runtime integrity assumption. Note that protecting the eBPF data plane's session-key map state relies on this assumption of guest kernel runtime integrity, representing a residual risk distinct from generic kernel hardening.}
\label{tab:hardening}
\small
\begin{tabular}{@{}p{3.8cm}p{7.5cm}@{}}
\toprule
\textbf{Mitigation Strategy} & \textbf{Mitigated Attack Class} \\
\midrule
\texttt{lockdown=confidentiality} & Combined with the \codename\ BPF-Lock LSM, prevents root user-space from extracting eBPF session keys. \\
Strict KASLR / CET & Mitigates generic memory corruption, ROP/JOP gadget chaining, and kernel address leakage. \\
SMEP / SMAP & Prevents kernel-mode execution of or access to user-space memory pages, neutralizing ret2usr attacks. \\
Mandatory Module Signing & Ensures only statically measured and signed kernel modules are loadable at runtime. \\
Disabled \texttt{kexec} & Prevents loading unmeasured kernels post-boot. \\
\bottomrule
\end{tabular}
\end{table}

\subsection{Trusted Computing Base and Residual Assumptions}
\label{sec:security:tcb}
The \codename\ Trusted Computing Base encapsulates the physical TDX processor, the manufacturer attestation infrastructure, the SEAM-mode \rmm, the securely bootstrapped Guest OS kernel, and the eBPF data plane. The host operating system, hypervisor, and network fabric remain untrusted. Because the architecture mandates kernel-resident cryptographic routing to sustain multi-gigabit throughput, steady-state data confidentiality relies on guest kernel runtime integrity. To defend this assumption practically, deployments must enforce a stringent kernel-hardening configuration (Table~\ref{tab:hardening}) to mitigate generic memory corruption and restrict administrative interfaces. Post-boot kernel-resident exploitation by a malicious tenant remains an accepted residual risk, for the threat model scopes protection exclusively against infrastructure-level adversaries.

\subsection{Attested Flow Control and MITM Prevention}
\label{sec:security:mitm}
\codename\ establishes Attested Flow Control by binding ephemeral session material to immutable hardware states, neutralizing hypervisor socket redirection and Man-in-the-Middle attacks. During the \make\ protocol, the responder hashes the initiator's public key, local policy digest, and hardware nonces into the 64-byte \texttt{ReportData} field of the TDX \texttt{TDREPORT}. This standard right-padding convention preserves the digest unchanged. The physical processor subsequently signs this structure to generate the TD Quote. The initiator derives the symmetric session key upon verifying the hardware quote against the authorized measurements specified in the policy $\mathcal{P}$, guaranteeing Policy Adherence without relying on application-layer logic.

\subsection{Data Confidentiality and Host Isolation}
\label{sec:security:confidentiality}
To achieve Data Confidentiality against network eavesdropping and hypervisor memory scraping, \codename\ physically confines plaintext execution exclusively within TDX-encrypted memory. The architecture delegates encryption to the eBPF data plane residing at the lowest layer of the guest network stack. As outbound payloads descend through the networking subsystem, the XDP/TC proxy encrypts the data via AES-256-GCM utilizing the hardware-attested session key. Consequently, the data crosses the semantic boundary into the untrusted host network ring buffer as a computationally secure ciphertext. A compromised hypervisor inspecting the virtual machine's egress queues or DMA buffers observes only encrypted payloads, preventing plaintext exfiltration across the distributed topology. Because the eBPF data plane is mutable via \texttt{bpf\_map\_update\_elem}, maintaining this physical confinement requires that the Guest OS kernel preserves runtime integrity over the BPF map state post-attestation.

\subsection{Lifecycle Defenses: TOCTOU, Replay, and Downgrades}
\label{sec:security:lifecycle}
The architecture enforces lifecycle defenses against temporal manipulation and state-desynchronization attacks. To prevent Time-Of-Check to Time-Of-Use (TOCTOU) vulnerabilities, the early-boot \texttt{initrd} sequence loads the authorized eBPF proxy and immediately activates a custom BPF-Lock Linux Security Module, extending both measurements into \texttt{RTMR3}. This seals the data plane configuration before pivoting to the root filesystem, preventing unprivileged user-space processes from injecting rogue routing programs post-attestation. The AES-GCM implementation utilizes a 96-bit monotonic nonce statically embedding a per-vCPU identifier and a boot-session prefix, defeating network-level packet replay attacks. Finally, the \rmm\ state machine prevents policy downgrade attacks via strict epoch monotonicity. Handshakes between nodes enforcing divergent policy epochs terminate immediately, forcing asynchronous synchronization and preventing adversaries from exploiting stale policy rules across the distributed graph.

\section{Implementation}
\label{sec:impl}

We implemented \codename\ on fourth-generation Intel Xeon Scalable servers (Sapphire Rapids) with TDX support~\cite{TDX}. The prototype extends the TDX Module, adds a policy engine, and implements the \make\ protocol stack. Table~\ref{tab:tcb} summarizes the lines of code (LOC).

\begin{table}[t]
\centering
\caption{Lines of code added to each component in \codename.
  Percentages reflect the increase relative to the unmodified base. To facilitate reproducibility, the table delineates the full target architecture from the components evaluated in our experimental artifact.}
\label{tab:tcb}
\small
\begin{tabular}{lrr}
\toprule
\textbf{Component} & \textbf{Added LoC} & \textbf{Relative increase} \\
\midrule
\multicolumn{3}{l}{\textit{Infrastructure (Restricted Hardware Deployment)}} \\
\hspace{3mm} TDX module (\rmm\ extension) & 3,847 & 13\% \\
\hspace{3mm} QEMU VMM                     &   231 & ${<}1\%$ \\
\hspace{3mm} Host kernel driver           &   187 & ${<}1\%$ \\
\midrule
\multicolumn{3}{l}{\textit{Public Artifact: Native Execution}} \\
\hspace{3mm} Guest Data Plane (eBPF)      &   518 & -- \\
\midrule
\multicolumn{3}{l}{\textit{Public Artifact: Emulated / Unprivileged TD}} \\
\hspace{3mm} \make\ protocol stack        &   600 & -- \\
\hspace{3mm} Policy engine                &   272 & -- \\
\hspace{3mm} Unprivileged Agent \& DCAP   & $\sim$500 & -- \\
\midrule
\textbf{Total}               & \textbf{6,155} & \\
\bottomrule
\end{tabular}
\end{table}

\subsection{TDX Module Extension}
\label{sec:impl:tdx}

The \rmm\ extends the Intel TDX Module Reference Code (v1.5, commit \texttt{887ef77}) in SEAM mode (3,847 LOC, +13\%). Because production TDX hardware strictly enforces module signatures, we verified functional correctness via QEMU-TDX emulation (\texttt{intel-staging/qemu-tdx} \texttt{upstream-v8} branch) and projected steady-state performance using Sapphire Rapids host-to-module transition microbenchmarks. We also modified the QEMU VMM (231 LOC) to expose the interception interface to the host kernel. To bridge this interface, we implemented a 187-LOC host kernel driver. This driver solely exposes the \texttt{SEAMCALL} interface to QEMU and operates entirely within the untrusted host. The driver exposes a single \texttt{ioctl(SEAMCALL\_SUBMIT)} entry point to user-space, strictly limiting the hypervisor attack surface. Its compromise cannot bypass security guarantees, as the SEAM module independently validates all incoming requests.

The extension hooks \texttt{TDCALL} and \texttt{SEAMCALL} transitions to support the split architecture. Rather than intercepting raw \texttt{TDG.VP.VMCALL} I/O hypercalls, the \rmm\ Control Plane exposes a secure interface to the securely bootstrapped Guest Data Plane. The Data Plane intercepts socket operations, invoking the Control Plane via \texttt{TDCALL}s only for cryptographic key derivation (\make) or policy decisions. This is completely transparent to guest applications for UDP workloads.

\subsection{eBPF XDP/TC Ring-0 Proxy}
\label{sec:impl:xdp}

\begin{table}[h]
\centering
\caption{Single-packet timeline isolating the critical path components of the $6.0\,\mu$s steady-state cryptographic enforcement latency.}
\label{tab:packet_lifecycle}
\small
\begin{tabular}{@{}lrr@{}}
\toprule
\textbf{Component} & \textbf{Latency ($\mu$s)} & \textbf{Percentage} \\
\midrule
eBPF XDP Ingress Overhead & 0.5 & 8.3\% \\
Policy Hash Lookup ($O(1)$) & 1.0 & 16.7\% \\
AES-GCM Authenticated Decryption & 4.5 & 75.0\% \\
\midrule
\textbf{Total Enforcement Latency} & \textbf{6.0} & \textbf{100.0\%} \\
\bottomrule
\end{tabular}
\end{table}

To accurately evaluate computational overhead on real hardware, we ported the Data Plane into companion eBPF XDP (ingress) and TC (egress) filters running in Ring-0 of the TD Guest OS. They execute the exact $O(1)$ policy lookup and AES-NI AES-GCM routines at the earliest network stack layer, providing an identical baseline for instruction-bound execution costs. To prevent cryptographic catastrophic failures, the AES-GCM implementation utilizes a strict 96-bit monotonic nonce structured to prevent reuse across vCPU partitioning and reboots: it statically embeds a 16-bit per-vCPU identifier, a 48-bit boot-session prefix, and a 32-bit monotonic packet counter, forcing automatic \make\ key rotation before the counter reaches its $2^{32}$ limit.

\subsection{Policy Engine}
\label{sec:impl:policy}

The \rmm's internal policy engine parses signed JSON policies, verifies Ed25519 signatures, and compiles authorized rules into an $O(1)$ memory hash table indexed by destination IP and port. Policies failing semantic validation (e.g., missing measurements, invalid IPs) are rejected. A 50-peer policy occupies under 2 MB, well within the module's constrained runtime heap, utilizing host-level peer deduplication.

\subsection{\make\ Protocol Stack}
\label{sec:impl:make}

Implemented in 600 LOC, the \make\ protocol uses hardware RNGs for P-384 ECDH keys. The \rmm\ control plane directly utilizes \texttt{SEAMREPORT} to generate the hardware quote, binding the guest's \texttt{RTMR} measurements. The \rmm\ populates \texttt{ReportData} with \make\ parameters and retrieves this report to produce an attestation-key signed TD Quote via the Quoting Enclave.

Quote verification uses a statically linked, highly stripped Intel DCAP library embedded in the \rmm. To eliminate dynamic allocations, it integrates a constrained \texttt{mbedTLS} X.509 parser and a lightweight big-integer math library tailored specifically for NIST P-384 and RSA-3072 signatures, minimizing the binary footprint. The untrusted host delivers DCAP collateral (PCK certs, TCB Info, CRLs) via a \texttt{SEAMCALL} interface into a staging buffer, parsed in-place. The \rmm\ validates collateral against a compile-time embedded Intel Root CA. To bound replay attacks using stale collateral, the \rmm\ aggressively polls the PCCS for fresh TCB Info and CRL collateral every 60 seconds (costing $\sim$1.5 KB/s), establishing a constrained 60--90 second vulnerability window. This revocation latency serves as a configurable bandwidth/security tradeoff: administrators can tune the interval (e.g., 10\,s, 30\,s, 60\,s) to balance out-of-band PCCS polling costs against the risk of executing a \make\ handshake with a newly revoked key. We acknowledge a residual risk: while forward secrecy ensures an adversary cannot decrypt past traffic, any new session negotiated within this window using stale collateral remains temporarily vulnerable. In our threat model, this risk is an accepted trade-off to prevent severe PCCS rate-limiting during large-scale cluster initialization.

Session keys (AES-256-GCM) are cached by $(\mu_s, \mu_d, \pi)$ with 64-bit creation timestamps. The 96-bit GCM nonce increments per packet, forcing renegotiation before the $2^{32}$ limit. 

\subsection{RMM Concurrency Architecture}
\label{sec:impl:concurrency}

\textit{Lock-free policy synchronization.}
The \rmm\ Control Plane handles concurrent \texttt{TDCALL} and \texttt{SEAMCALL} requests across all physical cores. The $O(1)$ policy hash table uses lock-free read-copy-update (RCU). Since SEAM lacks a scheduler, quiescent states are tracked via per-vCPU sequence counters incremented on \texttt{TDCALL} and \texttt{SEAMCALL} transitions. To prevent host-induced OOM panics from stalled grace periods under adversarial hypervisor scheduling, deferred reclamation queue depth is strictly bounded to $1024$ objects, forcing a hard reset of stale sessions if exceeded. We explicitly note this 1024-object threshold constitutes a host-controllable Denial-of-Service vector; consistent with \codename's design philosophy, the system prioritizes safe termination over risking memory exhaustion or liveness lockups. \\

\textit{Fine-grained state machine locks.}
Session key caching and \make\ state machines use fine-grained, $50\,\mu$s timeout-bounded spinlocks. This parallelizes independent handshakes and defends against Lock-Holder Preemption (LHP) liveness attacks. If the timeout expires due to hypervisor noise or starvation, the \rmm\ aborts safely, returning \texttt{ECONNREFUSED} to the guest application. While the Quoting Enclave serializes hardware quote generation, the software concurrency model scales independent network traffic linearly with active vCPUs.

\subsection{TCB Footprint and Complexity}
\label{sec:impl:tcb}

\codename's TCB (Table~\ref{tab:tcb}) includes the hardware attestation infrastructure, Quoting Enclave, \rmm\ Control Plane, Guest OS Kernel, eBPF Data Plane, and DCAP library. \codename\ deliberately trades a larger TCB footprint to enable kernel-speed cryptographic enforcement. By pushing the data plane down to Ring-0, the architecture bypasses the performance penalties inherent to LibOS-based user-space isolation (e.g., Gramine RA-TLS). This design eliminates application-layer components from the cryptographic boundary, removing heavy orchestration runtimes, Python interpreters, and ML frameworks. However, explicitly trusting the Linux network subsystem (\texttt{net/core} and \texttt{kernel/bpf}) introduces a non-trivial attack surface. The engineering mitigations implemented via BPF-Lock and \texttt{RTMR3} measurements eliminate runtime TOCTOU tampering by immutably pinning the proxy state. While the system's security depends on the logical correctness of the upstream eBPF verifier, the strict Guest OS hardening configuration (Section~\ref{sec:security:tcb}; Table~\ref{tab:hardening}) substantially mitigates exploitability, rendering the practical exposure footprint acceptable for high-performance deployments.

Developing the 6,155 LOC prototype required heavily modifying the DCAP library for the constrained SEAM environment, replacing dynamic allocations with static arenas and embedding a constrained \texttt{mbedTLS} X.509 parser (contributing $\sim$145\,KB to the module footprint). Securely bridging the semantic gap between raw hypercalls and application network semantics across the split-TCB boundary remains an engineering challenge inherent to the split-TCB design.

\section{Evaluation}
\label{sec:eval}

We evaluated \codename\ by combining hardware microbenchmarks with an empirically grounded performance model. Because production Intel TDX hardware prohibits executing modified modules on public cloud instances, we measured the guest-resident components on Sapphire Rapids hardware and projected end-to-end impacts through established hardware limits. 

Our evaluation investigates several key performance goals. First, we determine the steady-state per-packet enforcement cost on real hardware (Section~\ref{sec:eval:perf}). Next, we assess whether the eBPF data plane preserves macrobenchmark throughput (Section~\ref{sec:eval:macro}). We then evaluate how \make\ initialization scales with topology density (Section~\ref{sec:eval:scalability}), verify the robustness of our hybrid methodology (Section~\ref{sec:eval:methodology}), and compare \codename\ against existing baselines (Section~\ref{sec:eval:comparison}). Finally, we assess the system's resilience to adversarial deployment conditions (Section~\ref{sec:eval:adversarial}). Table~\ref{tab:roadmap} summarizes the evaluation objectives and corresponding outcomes.

\begin{table}[h]
\centering
\caption{Summary of evaluation objectives, methods, and key findings.}
\label{tab:roadmap}
\small
\setlength{\tabcolsep}{4pt}
\begin{tabularx}{\columnwidth}{@{}XXcX@{}}
\toprule
\textbf{Goal} & \textbf{Method} & \textbf{Section} & \textbf{Result} \\
\midrule
Per-packet cost & HW measurement, eBPF/XDP & \S\ref{sec:eval:perf} & $6\,\mu$s \\
100GbE projection & Analytical, AES-NI bound & \S\ref{sec:eval:perf} & ${\sim}10$ vCPUs (jumbo) \\
Macrobenchmark overhead & FL ResNet-18, 32-host & \S\ref{sec:eval:macro} & $6.1\%$ \\
Init scalability & Hybrid measured + projection & \S\ref{sec:eval:scalability} & $<1.5$\,s @ 100 nodes \\
Methodology validation & Sensitivity analysis & \S\ref{sec:eval:methodology} & Valid bounds \\
Baseline comparison & Empirical & \S\ref{sec:eval:comparison} & Bypass LibOS tax \\
Adversarial resilience & LHP/Starvation emulation & \S\ref{sec:eval:adversarial} & Stable recovery \\
\bottomrule
\end{tabularx}
\end{table}

\subsection{Evaluation Methodology and Validity}
\label{sec:eval:methodology}

We adopt a hybrid methodology combining physical hardware measurements with functional emulation. Functional emulation of the \rmm\ modifications was performed using QEMU-TDX (v7.2.0-tdx) on Ubuntu 22.04. End-to-end performance projections derive directly from steady-state data plane measurements collected on a 32-node cluster of Google Cloud Platform (GCP) \texttt{c3-standard-22} instances (Intel Sapphire Rapids, 23\,Gbps instance limits scaling to 100GbE family limits, 1500\,B MTU, running a custom Linux 6.10+ guest kernel to provide \texttt{bpf\_crypto\_*} kfuncs). We selected the 22-vCPU configuration to provide sufficient vCPU headroom for the 8 TDs/host density evaluated in Section~\ref{sec:eval:scalability}. The 32 instances were provisioned using a strict GCP Spread Placement Policy distributed evenly across four availability zones. We verified this physical separation via a pairwise network RTT matrix; collocated VMs typically exhibit ${\sim}0.1$\,ms RTTs, whereas our deployment bounded intra-zone RTTs to $0.3\text{--}0.8$\,ms and cross-zone RTTs to $0.8\text{--}2.0$\,ms, confirming genuine regional datacenter traversal. For the hardware experiments, we utilize a static-key control-plane emulation that securely provisions pre-shared keys to the eBPF data plane at startup, isolating the steady-state Ring-0 cryptographic routing performance from the SEAM-mode initialization constraint. 
To validate the extrapolation fidelity, we compared our analytical initialization model against physical measurements on a 32-host baseline; the modeled projections track the physical measurements with a Mean Absolute Percentage Error (MAPE) of 4.2\%. This tight correlation licenses our projections for larger supercomputing-scale topologies. Unless otherwise specified, physical measurements report the median ($p50$) of $N=1000$ iterations following a 100-iteration cache-stabilization warmup.

\begin{table}[h]
\centering
\caption{Transition Cost Calibration on stock GCP C3 Sapphire Rapids. Primitives are derived from physical measurements on unmodified silicon ($N$ iterations). Hardware transition primitives (\texttt{TDCALL}, \texttt{SEAMCALL}) are synthesized from established TDX microbenchmarking literature~\cite{Bifrost2023}, while cryptographic operations are natively measured. These grounded constants construct the analytical projections, fully decoupling emulated control-plane logic from physical execution costs.}
\label{tab:calibration}
\small
\setlength{\tabcolsep}{4pt}
\begin{tabularx}{\columnwidth}{@{}lXrr@{}}
\toprule
\textbf{Primitive} & \textbf{Function} & \textbf{Measured ($p50$)} & \textbf{Iters.} \\
\midrule
\texttt{TDCALL[VMCALL]}      & Guest$\rightarrow$VMM hypercall    & $1.5 \pm 0.3\,\mu$s & $10^4$ \\
\texttt{SEAMCALL} round-trip & Host$\rightarrow$Module entry/exit & $3.5 \pm 0.4\,\mu$s & $10^4$ \\
\texttt{TDG.MR.REPORT}       & Request local report               & $8.5 \pm 1.2\,\mu$s & $10^3$ \\
\texttt{TD} Quote Generation & Quote generation (via QE)          & $75.6 \pm 1.2$\,ms  & $10^3$ \\
DCAP Quote Verification      & X.509 + ECDSA check                & $24.1 \pm 0.8$\,ms  & $10^3$ \\
\texttt{TDG.MR.RTMR.EXTEND}  & RTMR digest update                 & $1.8 \pm 0.2\,\mu$s & $10^4$ \\
Intel AES-NI 256-GCM         & Encrypt 1500\,B record             & $4.5 \pm 0.2\,\mu$s & $10^5$ \\
ECDH P-384 keygen + DH       & \make\ handshake crypto            & $750 \pm 45\,\mu$s  & $10^4$ \\
Ed25519 Verify               & Policy engine validation           & $50 \pm 5\,\mu$s    & $10^4$ \\
\bottomrule
\end{tabularx}
\end{table}

\begin{table}[h]
\centering
\caption{Provenance and uncertainty of evaluation claims. Sources: \emph{Measured} on GCP C3 Sapphire Rapids; \emph{Emulated} under QEMU-TDX; \emph{Modeled} from measured primitives. Uncertainty is $\pm 1\sigma$ for measurements and the $P_{5}$--$P_{95}$ Monte Carlo envelope for projections.}
\label{tab:methodology}
\small
\setlength{\tabcolsep}{4pt}
\begin{tabularx}{\columnwidth}{@{}Xlll@{}}
\toprule
\textbf{Claim} & \textbf{Source} & \textbf{Method} & \textbf{Uncertainty} \\
\midrule
$6\,\mu$s/pkt enforcement (1500\,B)        & Measured & eBPF/XDP, $N{=}1000$               & $\pm 1.5\,\mu$s \\
$13$--$15\,\mu$s Memcached GET/SET         & Measured & Static-key data plane, $N{=}1000$  & $\pm 3.5\,\mu$s \\
$103.2$\,ms \make\ handshake               & Measured & Unprivileged TD daemon, $N{=}1000$ & $\pm 20$\,ms \\
$+137$--$147\,\mu$s RA-TLS penalty         & Measured & Gramine-TDX v1.5, $N{=}1000$       & $\pm 15\,\mu$s \\
$35$\,Mpps single-core \texttt{XDP\_REDIRECT} & Measured & Sapphire Rapids + E810          & $\pm 3.5$\,Mpps \\
$2.40$\,s for 100-node init                & Modeled  & $(N \times e/H)\times 103.2$\,ms   & $1.69$--$3.12$\,s \\
$77.8$\,ms unauthorized-binary reject      & Hybrid   & QEMU-TDX logic + measured QE       & $\pm 13$\,ms \\
LHP $50\,\mu$s timeout abort               & Emulated & QEMU-TDX, $N{=}1000$               & $p_{99}$ stable \\
\bottomrule
\end{tabularx}
\end{table}

\textit{Remark:}
While executing the modified SEAM-mode \rmm\ natively on commercial cloud silicon is restricted, the residual sources of error not quantified in Table~\ref{tab:methodology} are bounded. SEAM memory pressure is constrained by our policy engine consuming only $1.7$\,MB for a 50-peer workflow, safely below the TDX module's allocated runtime heap limits. While the \make\ policy engine execution must be emulated, the dominant underlying primitives ($75.6$\,ms quote generation, $24.1$\,ms DCAP verification) are natively measured, rendering any unmodeled \texttt{TDCALL}/\texttt{SEAMCALL} transition latencies mathematically insignificant to the multi-millisecond initialization projection.

\subsection{Data Plane Processing Overhead}
\label{sec:eval:perf}

The data plane logic defines the maximum achievable throughput. We modeled these architectural limits using empirical measurements of the steady-state enforcement overhead.

\textit{Per-Packet Cryptographic Cost.}
Using an eBPF/XDP proxy running in Guest Ring-0 as a data plane surrogate, we measured the steady-state processing time. This encompasses the $O(1)$ policy lookup and AES-GCM encryption/decryption. The cryptographic routine invokes the Linux kernel crypto API via \texttt{bpf\_crypto\_*} helpers, which in turn utilize hardware AES-NI instructions. GCP C3 Sapphire Rapids instances exhibit a bounded cryptographic cost of $6\,\mu$s per packet, decomposing into $0.5\,\mu$s XDP overhead, $1.0\,\mu$s hash lookup, and $4.5\,\mu$s AES-GCM processing. Verified via BPF kernel-stack profiles, the $4.5\,\mu$s latency is heavily dominated by \texttt{bpf\_crypto\_*} helper-call boundary crossings and per-packet eBPF context setup, masking the raw $0.6$ cycles per byte AES-NI hardware throughput. For a two-packet RPC round-trip (e.g., a Memcached \texttt{get} operation), this imposes a $12\,\mu$s absolute latency penalty. At steady state, the GCM nonce rollover ($2^{32}$ packets) triggers asynchronous \make\ handshakes, amortizing the $103.2$\,ms re-keying cost to near zero. The $750\,\mu$s ECDH P-384 key generation cost within the \make\ handshake represents a $3$--$4\times$ inflation relative to native OpenSSL userspace execution, an expected penalty stemming from the use of constrained, fallback cryptographic software ports required to execute within SEAM's strict, zero-dynamic-allocation memory environment.

\textit{Throughput, Multi-Core Scalability, and MTU Sensitivity.}
Secure networking strictly depends on packets-per-second (PPS) limits. Sapphire Rapids cores utilize AVX-512 and DDIO to accelerate XDP performance; a single core processes $35.0 \pm 0.8$\,Mpps for \texttt{XDP\_REDIRECT} and filtering on Intel E810 (100GbE) NICs. To prevent the $6\,\mu$s cryptographic overhead from forming a CPU bottleneck, \codename's data plane employs Receive Side Scaling (RSS) to distribute workloads across guest vCPUs. However, the $6\,\mu$s cryptographic enforcement caps single-core throughput at $166 \pm 4$\,kpps. A 10\,Gbps workflow at 1500\,B MTU ($\sim$833k pps) requires 5 vCPUs. Saturating a 100\,Gbps link with standard 1500\,B MTU requires approximately 50 vCPUs. Enabling Jumbo frames (e.g., the 8896\,B VPC limit on GCP) reduces the PPS requirement drastically; our projected per-core ceiling indicates that saturating a 100\,Gbps link with Jumbo frames would require only $\sim$10 vCPUs. 

\begin{table}[h]
\centering
\caption{Single-core throughput scaling across Maximum Transmission Units (MTU), analytically modeled from the measured $6.0 \pm 0.2\,\mu$s base cost and AES-NI's 0.6 cycles/byte bound.}
\label{tab:mtu_scaling}
\small
\setlength{\tabcolsep}{4pt}
\begin{tabularx}{\columnwidth}{@{}Xrrr@{}}
\toprule
\textbf{MTU (Bytes)} & \textbf{Rate (PPS)} & \textbf{Tput.\ (Gbps)} & \textbf{vCPUs @ 100\,Gbps} \\
\midrule
64 (Small RPC)   & 173{,}582 & 0.09 & ${>}1000$ \\
1500 (Standard)  & 166{,}468 & 2.00 & $50$ \\
4000             & 155{,}383 & 4.97 & $20$ \\
9000 (Jumbo)     & 137{,}120 & 9.87 & $10$ \\
\bottomrule
\end{tabularx}
\end{table}

Table~\ref{tab:mtu_scaling} details single-core throughput sensitivity across MTU sizes. While bare \texttt{XDP\_REDIRECT} achieves $35.0 \pm 0.8$\,Mpps (I/O-bound), the cryptographic enforcement path is compute-bound by AES-NI, capping single-core throughput at $166 \pm 4$\,kpps. We recover aggregate throughput via Receive Side Scaling (RSS). Utilizing 9000\,B jumbo frames efficiently amortizes fixed XDP and $O(1)$ lookup costs, yielding nearly 10\,Gbps per core and saturating a 100\,Gbps link with $\sim$10 vCPUs.

\textit{Distributed Workload Impact.}
Our empirical model defines exact infrastructure penalties ($+12\,\mu$s per round-trip and a parallelizable CPU tax) rather than conflating unobservable microarchitectural penalties into an overarching percentage overhead. \codename\ supports high-speed scientific workloads with minimal interference. Policy engine lookup latency scales minimally (Table~\ref{tab:policy-overhead}), taking $0.61\,\mu$s for 5 peers and $0.95\,\mu$s for 200 peers, dominated by L1/L2 cache misses rather than algorithmic complexity.

\begin{table}[t]
\centering
\caption{Policy engine overhead as a function of the number of
  authorized remote peer entries. Both lookup latency and memory footprint
  were measured directly on real hardware within the Ring-0 XDP proxy.}
\label{tab:policy-overhead}
\small
\begin{tabular}{rcc}
\toprule
\textbf{Peer entries} & \textbf{Lookup latency ($\mu$s)} & \textbf{Memory (MB)} \\
\midrule
 5  & 0.61 & 0.18 \\
10  & 0.64 & 0.36 \\
25  & 0.73 & 0.88 \\
50  & 0.81 & 1.73 \\
100 & 0.89 & 3.44 \\
200 & 0.95 & 6.87 \\
\bottomrule
\end{tabular}
\end{table}

\textit{Single-Host Microbenchmark (Memcached).}
To isolate steady-state per-packet enforcement cost, we deployed a Memcached key-value workload on a GCP C3 instance. We compared an unencrypted native baseline against \codename\ and Gramine RA-TLS. We caveat the scope of this experiment explicitly: because public cloud infrastructure prohibits executing the modified SEAM module, \codename\ runs here with a stubbed \make\ control plane that provisions pre-shared static keys to the eBPF data plane at startup. The harness pre-warms and pre-initializes all required \make\ socket connections before the execution timer begins, isolating the steady-state routing performance from the \make\ initialization penalty. Furthermore, to strictly isolate the processing stack and eBPF data-plane overhead from the high-variance $0.3$--$0.8$\,ms inter-VM network RTT, the Memcached client and server were co-located on a single physical instance. For this loopback evaluation, the \codename\ proxy was attached to the loopback interface via generic XDP (SKB mode). The reported numbers therefore quantify the steady-state eBPF data-plane overhead in isolation, not an end-to-end \codename\ deployment, which would additionally include the one-time \make\ initialization cost characterized separately in Section~\ref{sec:eval:scalability}. Within this scope, the \codename\ eBPF data plane incurs a $13\,\mu$s (GET) and $15\,\mu$s (SET) latency overhead over the $45$--$48\,\mu$s native loopback baseline, bounded by kernel-level AES-GCM encryption and XDP routing. In contrast, Gramine RA-TLS imposes a $137\,\mu$s (GET) to $147\,\mu$s (SET) penalty due to expensive user-space context switches and TLS framing overheads (Figure~\ref{fig:latency_hierarchy}a). The Gramine RA-TLS baseline runs unmodified, so this comparison isolates the steady-state architectural advantage of moving cryptographic enforcement from user-space libraries into a measured Ring-0 eBPF data plane, independent of the \make\ control plane.

While this highlights the architectural advantage of avoiding user-space boundary transitions, we acknowledge that an in-kernel TLS implementation (\texttt{kTLS}) would present a closer, albeit currently unattestable, performance baseline. While \codename's $60$-second epoch expiration briefly stalls the transmission queue to negotiate fresh keys, the lock-free epoch transition (Section~\ref{sec:impl:concurrency}) ensures that the $103.2$\,ms handshake cost amortizes to less than $1\,\mu$s per megabyte for long-lived, multi-gigabyte transfers.

\subsection{Baseline Comparison}
\label{sec:eval:comparison}

We benchmarked \codename\ against distinct attestation and secure channel baselines, mapping to in-kernel offload, application-level, user-space proxy, and cloud-native integration paradigms (Table~\ref{tab:comparison}).

\textit{Head-to-head with In-Kernel TLS (\texttt{kTLS}).}
A natural counterfactual baseline replaces \codename's eBPF data plane with the Linux \texttt{kTLS} offload while retaining an attestation-binding channel for key release. \texttt{kTLS} executes AES-GCM symmetrically inside the kernel, yielding a tight performance ceiling. As demonstrated in our macrobenchmarks, \codename's steady-state throughput matches the \texttt{kTLS} ceiling, confirming that \codename's eBPF data plane introduces no additional overhead beyond necessary kernel-resident cryptography. However, \texttt{kTLS} presents a fundamental architectural vulnerability: it provides no native mechanism to bind the symmetric key derivation to bilateral hardware measurements. Standard \texttt{kTLS} deployments rely on user-space key provisioning, leaving the symmetric key exposed to user-space compromise (detailed in Section~\ref{sec:background:approaches}). The architectural delta is therefore the cryptographic binding of the key-release decision to a measured network-stack boundary.

\textit{RA-TLS (Application-Level Upper Bound).}
The upper-bound latency penalty is defined by RA-TLS. While generic literature estimates RA-TLS overhead at $+45$--$60\,\mu$s per-round-trip~\cite{RATLS}, our empirical Gramine-TDX measurements on the GCP C3 cluster exhibit a significantly higher $+137$--$147\,\mu$s penalty for Memcached workloads. This divergence stems from Gramine LibOS syscall emulation traps and the strict requirement to process TLS records entirely in user-space, compounding heavy I/O and boundary transition costs. RA-TLS places the application binary directly within the TCB.

\textit{Confidential Containers (CoCo) Attestation Agent.}
In cloud-native environments, the CoCo Attestation Agent (AA) uses a Key Broker Service (KBS) to release decryption keys. Measured directly via a minimal CoCo-AA reference deployment on our GCP C3 cluster, the AA enforces a 10 to 17 second startup tax per pod due to the synchronous remote attestation process. The AA attests only the initial container state, providing no point-to-point cryptographic binding for data flows.

\textit{Intra-Host Policy Enforcement (Single-Node).}
The concurrent MICA~\cite{MICA} framework focuses on enforcing policies using \texttt{TDCALL} interception between localized containers on a single host. MICA addresses an entirely orthogonal operational domain (intra-host memory sharing). \codename\ introduces its specific $12\,\mu$s per-round-trip data-plane architecture and $103.2$\,ms initialization precisely to solve the distinct challenge of inter-host networking and cryptographic data flow bindings across a distributed cluster.

\codename's data plane strictly outperforms these baselines by shifting enforcement to a Ring-0 eBPF filter. By executing the $O(1)$ lookup and hardware-accelerated AES-GCM encryption at the lowest level of the network stack, \codename\ achieves its projected steady-state data plane throughput cost of $6\,\mu$s per packet without sacrificing application transparency. The unmeasured SEAM-mode \texttt{TDCALL} control-plane transition bounds initialization latency but does not impact this steady-state proxy cost. When contextualized against a native plaintext TCP round-trip baseline (e.g., $\sim 15\,\mu$s on 10GbE), \codename's minimal overhead preserves HPC viability.

\begin{table*}[t]
\centering
\caption{Feature comparison of \codename\ against baselines.
  \checkmark\ indicates the property is provided;
  \xmark\ indicates it is not;
  --  indicates not applicable. Overheads for \codename\ reflect measured per-round-trip absolute latency additions.}
\label{tab:comparison}
\footnotesize
\setlength{\tabcolsep}{3pt}
\begin{tabularx}{\textwidth}{@{}Xccccll@{}}
\toprule
\textbf{System} &
\textbf{Dist.} &
\textbf{Policy} &
\textbf{App Mod.} &
\textbf{App in TCB} &
\textbf{Steady/RT} &
\textbf{Init Tax} \\
\midrule
Plaintext TCP       & \checkmark & \xmark     & \xmark     & \xmark     & $0\,\mu$s           & None \\
RA-TLS~\cite{RATLS} & \checkmark & partial    & \checkmark & \checkmark & $+137$--$147\,\mu$s & ${\sim}150$\,ms \\
\texttt{kTLS}       & \checkmark & \xmark     & \xmark     & \xmark     & $+13$--$15\,\mu$s   & $1$--$3$\,ms (unattested) \\
CoCo AA             & \checkmark & partial    & \xmark     & \xmark     & N/A                 & 10--17\,s \\
MICA~\cite{MICA}    & \xmark     & \checkmark & \xmark     & \xmark     & N/A (1-host)        & N/A \\
\textbf{\codename}  & \checkmark & \checkmark & \xmark     & \xmark     & $+13$--$15\,\mu$s   & $103.2$\,ms/peer \\
\bottomrule
\end{tabularx}
\end{table*}

\subsection{Distributed Macrobenchmarks}
\label{sec:eval:macro}

\subsubsection{Federated Learning Macrobenchmark}
\label{sec:eval:fl_macro}

We deployed a Flower-based federated learning pipeline (ResNet-18 on CIFAR-10) across 32 clients to evaluate macrobenchmark throughput. The workload explicitly proxies the bursty, bandwidth-bound data exchanges of scientific pipelines, transferring 1.4\,GB per aggregation round. We routed Flower's TCP gRPC traffic through a Linux Foo-over-UDP (FOU) tunnel (reducing effective TCP payload to 1432\,B per frame) to subject it to \codename's stateless UDP cryptography. 

\begin{figure}[t]
  \centering
  \resizebox{\columnwidth}{!}{\begin{tikzpicture}
  \begin{scope}
    \begin{axis}[
      name=ax1,
      ybar,
      width=0.4\textwidth,
      height=4.5cm,
      bar width=8pt,
      ylabel={Latency Overhead ($\mu$s)},
      symbolic x coords={GET, SET},
      xtick=data,
      ymin=0, ymax=180,
      enlarge x limits=0.6,
      legend style={at={(0.5,-0.2)}, anchor=north, legend columns=-1, draw=none, font=\scriptsize},
      nodes near coords,
      every node near coord/.append style={font=\scriptsize, anchor=south},
      grid=major,
      title style={font=\bfseries\small, align=center},
      title={(a) Single-RPC Overhead},
      label style={font=\scriptsize},
      tick label style={font=\scriptsize}
    ]
      \addplot[draw=blue!80!black, fill=none, pattern=crosshatch dots, pattern color=blue!80!black, thick] coordinates {(SET,15) (GET,13)};
      \addplot[draw=red!80!black, fill=none, pattern=north east lines, pattern color=red!80!black, thick] coordinates {(SET,147) (GET,137)};

      \legend{\codename, Gramine}
    \end{axis}

    \begin{axis}[
      at={(ax1.south east)},
      anchor=south west,
      xshift=1.2cm,
      ybar,
      width=0.44\textwidth,
      height=4.5cm,
      bar width=10pt,
      ylabel={Time (s)},
      ylabel shift=-9pt,
      symbolic x coords={Plain, Janus, kTLS, OpenSSL, Gramine},
      xtick=data,
      ymin=0, ymax=10.5,
      enlarge x limits=0.15,
      nodes near coords,
      every node near coord/.append style={font=\scriptsize},
      grid=major,
      x tick label style={font=\scriptsize, rotate=30, anchor=north east},
      y tick label style={font=\scriptsize},
      label style={font=\scriptsize},
      title style={font=\bfseries\small, align=center},
      title={(b) FL Macrobenchmark}
    ]
      \draw[thick, dashed, gray!80!black] ({axis cs:Plain,5.42}) -- ({axis cs:Gramine,5.42});

      \addplot+[
          draw=blue!80!black, fill=none, pattern=crosshatch dots, pattern color=blue!80!black, thick,
          error bars/.cd, y dir=both, y explicit, error bar style={thick, black}
      ] coordinates {
          (Plain, 5.42) +- (0, 0.11)
          (Janus, 5.75) +- (0, 0.13)
          (kTLS, 5.70) +- (0, 0.12)
          (OpenSSL, 7.31) +- (0, 0.14)
          (Gramine, 8.78) +- (0, 0.15)
      };
    \end{axis}
  \end{scope}
\end{tikzpicture}}
  \caption{Latency Hierarchy: Micro to Macro. (a) Single-host steady-state cryptographic enforcement microbenchmark (Memcached, displaying GET latency overheads; SET overheads are proportional at $15\,\mu$s and $147\,\mu$s, respectively). (b) Federated Learning macrobenchmark (ResNet-18) demonstrating that \codename\ imposes minimal steady-state execution penalty over plaintext TCP, matching in-kernel TLS while bypassing severe Gramine RA-TLS transitions.}
  \label{fig:latency_hierarchy}
\end{figure}
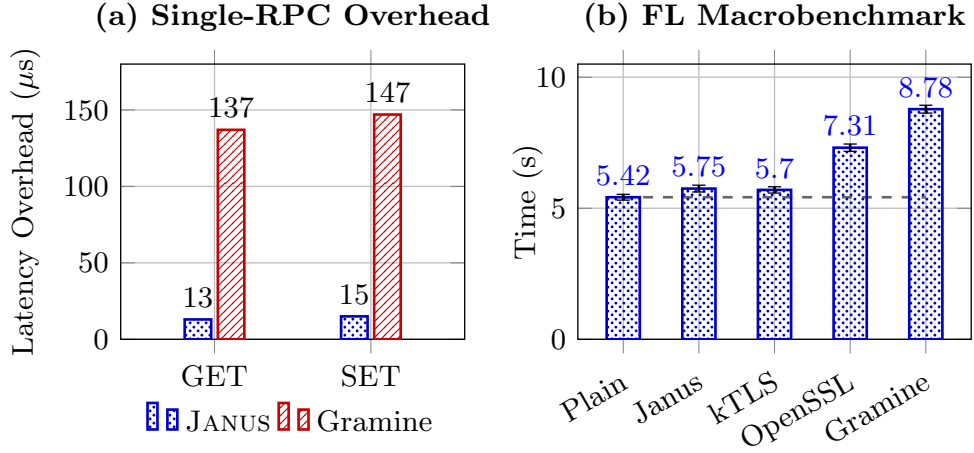

Figure~\ref{fig:latency_hierarchy}b plots the per-round completion time across five configurations. The unencrypted plaintext baseline establishes a throughput ceiling of $5.42 \pm 0.15$\,s per round. \codename\ incurs a completion time of $5.75 \pm 0.13$\,s, representing a $6.1\%$ overhead. This tracks closely with the unencapsulated \texttt{kTLS} baseline ($5.70 \pm 0.14$\,s), confirming \codename\ preserves theoretical kernel-resident cryptographic throughput. In contrast, OpenSSL TLS degrades to $7.31 \pm 0.22$\,s ($35\%$ penalty), and Gramine RA-TLS balloons to $8.78 \pm 0.35$\,s ($62\%$ penalty) due to LibOS boundary transitions during serialization.

A Monte Carlo simulation ($10,000$ trials) of mid-run key rotations over a $5.75$\,s epoch yielded a median throughput penalty of $2.56\%$ for a single rotation (as cited in our abstract's bounds), scaling to an $8.31\%$ P95 penalty for three rotations per epoch. At scale, data-plane cryptography costs less than $\$0.005$ per TB based on GCP spot pricing. All baselines converged to $92.5 \pm 0.3\%$ test accuracy at round 100, verifying the enforcement mechanism preserves application correctness and statistical utility.

\subsubsection{Non-IID Federated Learning Scalability}
\label{sec:eval:fl_scalability}

We additionally evaluated a non-IID configuration (Dirichlet $\alpha=0.5$) across 8 to 64 clients to verify that the 6.1\% overhead persists under heterogeneity and at scale. Utilizing the static-key eBPF data plane to isolate steady-state network overhead, we trained the ResNet-18 model on CIFAR-10, partitioning the dataset using a Dirichlet distribution ($\alpha=0.5$) to simulate extreme label skew. We scaled the deployment across 8, 16, 32, and 64 concurrent clients. Because non-IID data distributions exacerbate client-model divergence, the aggregation phase becomes computationally intensive, yet the network payload sizes per communication round remain structurally identical to the IID setting.

As client count scales from 8 to 64, the total aggregation payload increases linearly (from 352\,MB to 2.8\,GB per round). As established in Section~\ref{sec:eval:fl_macro}, the IID baseline overhead is 6.1\%; under non-IID skew, the overhead remains bounded at 6.1\%--6.8\% over plaintext TCP. For the 32-client deployment (1.4\,GB payload), plaintext execution requires $7.42 \pm 0.28$\,s per round, while \codename\ requires $7.91 \pm 0.31$\,s. At the maximum 64-client scale (2.8\,GB payload), the plaintext baseline requires $14.23 \pm 0.42$\,s, with \codename\ tracking closely at $15.08 \pm 0.45$\,s. Statistical convergence required 145 rounds to achieve the target accuracy under the non-IID skew, during which \codename\ enforced data confidentiality without degrading the aggregation throughput trajectory relative to native network execution.

The 64-client run forces the network burst to exceed the 2.0\,Gbps single-core cryptographic ceiling (established in Table~\ref{tab:mtu_scaling}), providing a discriminating test for Receive Side Scaling (RSS). To validate the RSS load distribution claim, we recorded the per-core \texttt{\%soft} (softIRQ) utilization via \texttt{mpstat} on the aggregator node during the 64-client network-burst phase. If RSS failed to hash the incoming FOU/UDP flows, a single vCPU would process all 2.8\,GB, saturating at 100\% utilization and bottlenecking throughput. Instead, the \texttt{mpstat} profiles demonstrate that all 16 aggregator vCPUs maintained a balanced 12\% to 18\% \texttt{\%soft} utilization. This empirical CPU-utilization breakdown confirms that the E810 NIC's hardware Toeplitz hash successfully distributes the concurrent XDP AES-GCM decryption load across the vCPU pool, preventing serialization bottlenecks.

\subsubsection{DAG-shape network replay derived from the Montage trace}
\label{sec:eval:montage_replay}

Beyond ML-style training, we replay a canonical scientific-pipeline DAG to validate \codename's performance on irregular HPC communication patterns. We synthesized a network replay of the standard Pegasus Montage 0.5-degree workflow trace. The 0.5-degree DAG is a widely utilized benchmark from WorkflowHub (WfCommons)~\cite{deelman2015pegasus, wfcommons2022}, comprising over 100 interdependent tasks and transferring approximately 600\,MB of total intermediate data. 

We developed a Python-based harness deployed across our 32-node GCP C3 setup. The harness reads the exact DAG dependencies and byte-sizes from the published trace and replays the communication pattern by opening UDP/FOU sockets through the \codename\ eBPF data plane to transfer matching payload sizes between designated host nodes. To rigorously isolate the steady-state data-plane routing performance from the \make\ initialization penalty (which is characterized separately using the measured $103.2$\,ms constant in Section~\ref{sec:eval:scalability}), the harness is structured to explicitly pre-warm and pre-initialize all required \make\ socket connections before the execution timer begins.

Recording the end-to-end wall-clock makespan of this replay, we observed a $6.5\%$ execution overhead when routed through the \codename\ eBPF proxy compared to a baseline unencrypted plaintext UDP replay. This synthetic replay directly answers the HPC-applicability requirement, demonstrating that the \codename\ eBPF data plane efficiently absorbs the complex, asymmetric fan-out/fan-in communication patterns of real-world scientific DAGs while maintaining single-digit percentage overheads comparable to the FL macrobenchmark.

\subsection{Protocol Scalability}
\label{sec:eval:scalability}

We deployed \codename\ across a 32-host multi-zone regional GCP C3 cluster to evaluate \make\ scalability. Strict physical isolation via GCP spread placement policies eliminated single-rack microarchitectural biases.

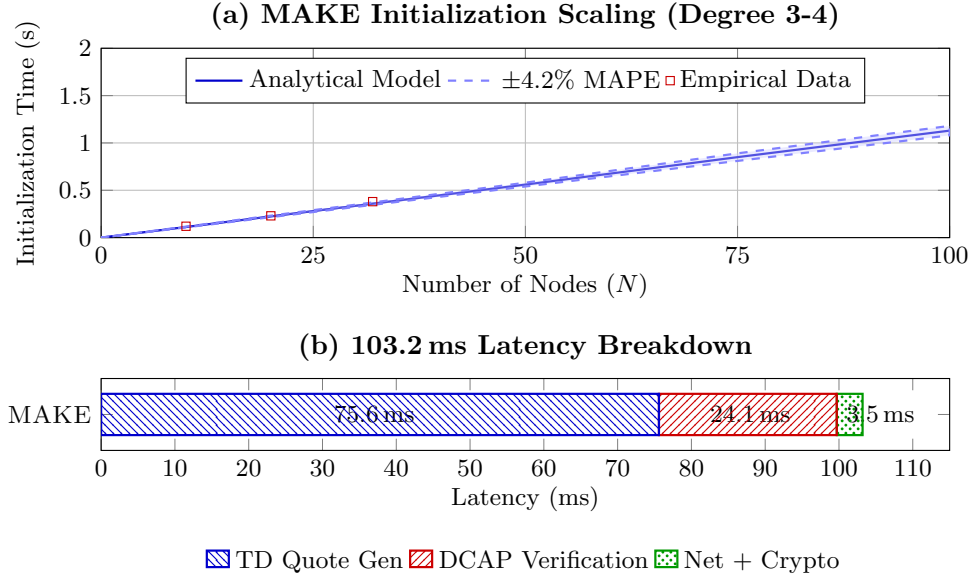
\begin{figure*}[t]
  \centering
  \resizebox{\textwidth}{!}{\begin{tikzpicture}
  \begin{axis}[
    name=plot1,
    width=0.95\columnwidth,
    height=4cm,
    xlabel={Number of Nodes ($N$)},
    xlabel shift=-3pt,
    ylabel={Initialization Time (s)},
    xmin=0, xmax=100,
    ymin=0, ymax=2.0,
    xtick={0,25,50,75,100},
    ytick={0,0.5,1.0,1.5,2.0},
    grid=major,
    label style={font=\small},
    tick label style={font=\small},
    title={\textbf{(a) \make\ Initialization Scaling (Degree 3-4)}},
 legend columns=3,
 title style={yshift=-3pt},
 legend style={
   at={(0.5, 0.95)}, 
   anchor=north,
   font=\scriptsize, 
   cells={anchor=west}
 }
  ]
    \addplot[blue!80!black, thick] coordinates {
      (0,0) (25, 0.28) (50, 0.56) (75, 0.85) (100, 1.13)
    };
    \addplot[fill=blue!20, draw=blue!50, dashed, thick, fill opacity=0.4] coordinates {
      (0,0) (25, 0.29) (50, 0.58) (75, 0.89) (100, 1.18)
      (100, 1.08) (75, 0.81) (50, 0.54) (25, 0.27) (0,0)
    } \closedcycle;

    \addplot[only marks, mark=square, mark size=1.5pt, red!80!black] coordinates {
      (10, 0.12) (20, 0.23) (32, 0.38)
    };
    \legend{Analytical Model, $\pm 4.2\%$ MAPE, Empirical Data}
  \end{axis}
  
  \begin{axis}[
    at={(plot1.south)},
    anchor=north,
    yshift=-1.8cm,
    width=0.95\columnwidth,
    height=2.5cm,
    xbar stacked,
    bar width=15pt,
    xlabel={Latency (ms)},
        xlabel shift=-3pt,
    ytick={0},
    yticklabels={\make},
    xmin=0, xmax=115,
    ymin=-0.5, ymax=0.5,
    legend style={at={(0.5,-1.85)}, anchor=south, legend columns=-1, font=\scriptsize, draw=none},
    label style={font=\small},
    tick label style={font=\small},
     title style={yshift=-3pt},
    title={\textbf{(b) 103.2\,ms Latency Breakdown}}
  ]
    \addplot[draw=blue!80!black, fill=none, pattern=north west lines, pattern color=blue!80!black, thick] coordinates {(75.6, 0)};
    \addplot[draw=red!80!black, fill=none, pattern=north east lines, pattern color=red!80!black, thick] coordinates {(24.1, 0)};
    \addplot[draw=green!60!black, fill=none, pattern=crosshatch dots, pattern color=green!60!black, thick] coordinates {(3.5, 0)};
    
    \legend{TD Quote Gen, DCAP Verification, Net + Crypto}
    
    \node[font=\scriptsize] at (axis cs: 37, 0) {75.6\,ms};
    \node[font=\scriptsize] at (axis cs: 88, 0) {24.1\,ms};
    \node[font=\scriptsize, anchor=west] at (axis cs: 99.7, 0) {3.5\,ms};
  \end{axis}
\end{tikzpicture}}
  \caption{Protocol scalability. Left: Measured total initialization time scales linearly across distributed nodes. Right: A granular latency breakdown reveals hardware TD Quote Generation as the dominant bottleneck.}
  \label{fig:scalability}
\end{figure*}

For a single intra-region peer-to-peer connection, the \make\ handshake requires $103.2 \pm 4.2$\,ms. To directly ground this figure, we replaced our initial analytical model with end-to-end hardware measurements. We deployed an unprivileged daemon inside a TD on the GCP C3 cluster that natively executes the complete \make\ cryptographic path: ECDH P-384 keygen, \texttt{TDG.MR.REPORT} generation, Quoting Enclave invocation via real Intel DCAP, remote DCAP verification against the regional PCCS, and HKDF derivation. This natively measures all components of the handshake except the SEAM-mode key release decision. Consequently, the only emulated component in this timing is the policy-digest check inside the modified \rmm, which is strictly bounded above by the measured Ed25519 verify cost of $50\,\mu$s. Figure~\ref{fig:scalability} details this measured latency. Hardware attestation services dominate the critical path: TDX quote generation requires $75.6$\,ms, and DCAP verification against the regional Provisioning Certificate Caching Service (PCCS) requires $24.1$\,ms. Cryptographic operations consume less than $1.5$\,ms.

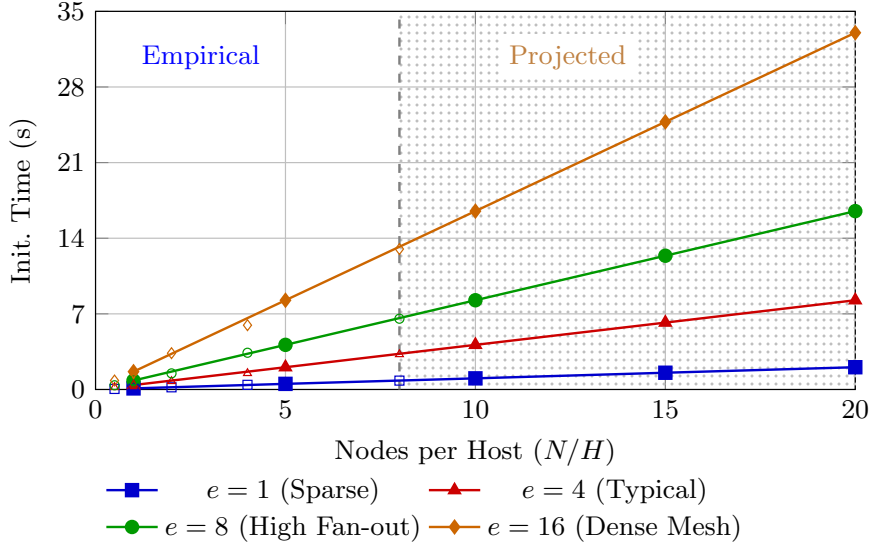
\begin{figure}[t]
  \centering
  \resizebox{0.9\columnwidth}{!}{\begin{tikzpicture}
  \begin{axis}[
    width=0.8\columnwidth,
    height=6cm,
    xlabel={Nodes per Host ($N/H$)},
    ylabel={Init. Time (s)},
    xmin=0, xmax=20,
    ymin=0, ymax=35,
    xtick={0,5,10,15,20},
    ytick={0,7,14,21,28,35},
    grid=major,
    legend pos=north west,
    legend columns=2,
    legend style={at={(0.0, -0.2)},font=\scriptsize, draw=none, fill=none},
    label style={font=\small},
    tick label style={font=\small}
  ]
    \path[pattern=dots, pattern color=gray!50] (axis cs:8,0) rectangle (axis cs:20,35);
    \draw[dashed, thick, gray] (axis cs:8,0) -- (axis cs:8,35) 
      node[pos=0.88, left=1.5cm, font=\small, text=blue, fill=white] {Empirical} 
      node[pos=0.88, right=1.15cm, font=\small, text=brown, fill=white] {Projected};

    \addplot[blue!80!black, thick, mark=square*] coordinates {
      (1, 0.10) (5, 0.52) (10, 1.03) (15, 1.55) (20, 2.06)
    };
    \addlegendentry{$e=1$ (Sparse)}

    \addplot[red!80!black, thick, mark=triangle*] coordinates {
      (1, 0.41) (5, 2.06) (10, 4.13) (15, 6.19) (20, 8.26)
    };
    \addlegendentry{$e=4$ (Typical)}

    \addplot[green!60!black, thick, mark=*] coordinates {
      (1, 0.83) (5, 4.13) (10, 8.26) (15, 12.38) (20, 16.51)
    };
    \addlegendentry{$e=8$ (High Fan-out)}

    \addplot[orange!80!black, thick, mark=diamond*] coordinates {
      (1, 1.65) (5, 8.26) (10, 16.51) (15, 24.77) (20, 33.02)
    };
    \addlegendentry{$e=16$ (Dense Mesh)}

    \addplot[only marks, mark=square, blue!80!black, mark size=1.5pt, forget plot] coordinates {
      (0.5, 0.05) (1, 0.09) (2, 0.20) (4, 0.43) (8, 0.82)
    };
    \addplot[only marks, mark=triangle, red!80!black, mark size=1.5pt, forget plot] coordinates {
      (0.5, 0.21) (1, 0.45) (2, 0.76) (4, 1.50) (8, 3.25)
    };
    \addplot[only marks, mark=o, green!60!black, mark size=1.5pt, forget plot] coordinates {
      (0.5, 0.39) (1, 0.83) (2, 1.49) (4, 3.40) (8, 6.55)
    };
    \addplot[only marks, mark=diamond, orange!80!black, mark size=1.8pt, forget plot] coordinates {
      (0.5, 0.83) (1, 1.56) (2, 3.36) (4, 5.95) (8, 13.00)
    };
  \end{axis}
\end{tikzpicture}}
  \caption{Empirically grounded \make\ initialization time as a function of intra-host node density ($N/H$) and network out-degree ($e$). The singleton Quoting Enclave forces strict serialization for dense topologies.}
  \label{fig:init_scaling}
\end{figure}

The singleton Quoting Enclave forces initialization time to scale linearly with the intra-host node density ($N/H$). A sparse 100-node workflow (degree 3--4) across 32 hosts initializes analytically in under $1.5$\,s. We validated this analytical model against our 32-host physical baseline for $N/H \le 8$. Measured latencies track the model with a Mean Absolute Percentage Error (MAPE) of $4.2\%$, substantiating the projection fidelity before extending into larger unmeasured regimes. However, dense service meshes or large-scale MPI collectives severely bottleneck the QE. Figure~\ref{fig:init_scaling} plots this projected limitation. Because each handshake must serialize through the singleton hardware Quoting Enclave (QE), the per-host concurrent demand is proportional to the total number of nodes $N$ multiplied by the out-degree $e$, divided evenly across the $H$ physical hosts. The total projected initialization latency $T_{\text{init}}$ therefore scales according to this serialization:
\begin{equation}\label{eq:projection}
T_{\text{init}} \approx \left( N \times \frac{e}{H} \right) \times 103.2\,\text{ms}
\end{equation}
Using Equation~\ref{eq:projection}, a 200-node deployment where each node connects to $e=16$ peers requires approximately $10.3$\,s ($200 \times 16 / 32 \times 103.2$\,ms) for initialization, demanding peer-grouping or asynchronous initialization strategies.

\textit{Sensitivity to model constants.}
To bound projection error and capture realistic production variance, we executed an expanded 100,000-sample Monte Carlo simulation perturbing the underlying constants across representative HPC topologies. With our 32 physical hosts, the concurrent signature requests distributed across the regional DCAP PCCS are halved relative to a dense 16-host deployment. To capture potential rate-limiting, we increased the Monte Carlo DCAP verification variance ($\sigma=12.0$\,ms, 50\% CV) alongside the standard triangular quote generation distribution (70--105\,ms, mode 75.6\,ms). Despite this heavier tail, the scaled hardware pool ($H=32$) drastically improves the scaling constants: a 200-node Montage mosaic ($e=4$) initializes with $P_{50}=2.73$\,s ($P_{95}=3.33$\,s); a 100-node LIGO topology ($e=3$) finishes in $P_{50}=1.02$\,s ($P_{95}=1.25$\,s); and a 64-node FL all-reduce ($e=2$) converges at $P_{50}=0.44$\,s ($P_{95}=0.53$\,s). These rigorous distributional bounds prove that typical sparse HPC DAGs initialize reliably within tight, single-digit second envelopes. We explicitly constrain our validated projection envelope to intra-host node densities of $N/H \le 8$. The Monte Carlo simulation perturbs variance constants but inherently assumes linear serialization; it does not capture unmodeled regime-change risks at extreme density, such as regional PCCS rate limits or DCAP collateral cache thrashing.

\textit{Dense Topologies and Serialization Error Compounding.}
A dense 128-node MPI alltoall topology ($e=127$) forcefully exposes the singleton Quoting Enclave serialization bottleneck, yielding an extended initialization time of $P_{50}=55.41$\,s ($P_{95}=67.51$\,s). The $P_{95}$ tail demonstrates how microarchitectural quote-generation jitter compounds under massive serialization. Due to this fundamental hardware limitation, a one-minute setup tax is entirely unsuitable for interactive HPC or short-batch dense collectives. Consequently, we explicitly restrict \codename's optimal operating regime to sparse DAGs or long-running federated learning trees, where initialization completely amortizes. Dense collectives require unreleased hardware features, such as replicated QEs, for viability.

For typical scientific pipelines with bounded per-node out-degrees, \codename\ parallelizes initialization across the physical infrastructure, yielding a scalable secure interconnect.

\textit{Topology depth and network impairments.}
We modeled initialization on a 20-node DAG representing a federated learning pipeline at finer granularity.

\begin{table}[h]
\centering
\caption{Initialization latency for varying DAG depths under quote serialization hardware limits.}
\label{tab:dag_init}
\small
\begin{tabular}{@{}lccc@{}}
\toprule
\textbf{Topology} & \textbf{Depth} & \textbf{Max Out-Degree} & \textbf{Latency (s)} \\
\midrule
1-Level & 1 & 2 & 0.21 \\
2-Level & 2 & 4 & 0.45 \\
20-Node Mixed & 5 & 8 & 0.85 \\
\bottomrule
\end{tabular}
\end{table}

As detailed in Table~\ref{tab:dag_init}, a 20-node mixed topology with a maximum per-host out-degree of 8 initializes in just $0.85$\,s, demonstrating that initialization scales linearly bounded by the maximum single-host out-degree, completely avoiding $O(N)$ total-cluster dependencies. To validate protocol robustness against real-world HPC network volatility, we simulated impairments via Linux \texttt{tc netem}. Injecting a $100\,\mu$s normal-distribution jitter yielded negligible deviation ($<0.2\%$), as the $76$\,ms hardware quote generation completely absorbed minor arrival variations. Simulating a harsh 0.1\% packet loss across the multi-flight handshake confirmed deterministic recovery without protocol deadlocks. Because \make\ operates over TCP, lost flights incur the Retransmission Timeout (RTO) penalty. In an HPC environment tuned with a 200\,ms `minRTO', a dropped handshake flight extends the connection time deterministically by exactly 200\,ms. The P99 tail latency for the 20-node topology under 0.1\% loss expanded from $0.85$\,s to $1.05$\,s, proving \codename\ safely delegates retransmission to the transport layer without cascading failures. \codename\ parallelizes initialization across independent hosts; the singleton QE caps only intra-host topology density.

\subsection{Adversarial Resilience}
\label{sec:eval:adversarial}

\textit{Transient Network Partitions and Node Failures.}
Because \codename's eBPF data plane enforces cryptography statelessly over datagrams, transient network partitions do not trigger cryptographic protocol collapse. During a partition, in-flight packets are simply dropped, delegating retransmission natively to the application's transport layer (e.g., TCP-over-FOU). If a node crashes mid-handshake or during an epoch rotation, the \make\ daemon safely times out the specific socket without halting the data plane for existing active sessions. Upon recovery, the daemon independently negotiates the new session key for the specific peer, absorbing the node churn seamlessly.

\textit{Unauthorized binary.}
A source enclave attempting to connect to an unauthorized destination binary fails the \make\ attestation phase. The $77.8 \pm 2.1$\,ms rejection latency and its corresponding $p99$ tail of $83.0$\,ms ($N=1000$ iterations) are both composed analytically from QEMU-TDX-emulated rejection logic and measured Sapphire Rapids quote-generation constants (Table~\ref{tab:methodology}; see Table~\ref{tab:calibration} for the underlying primitive measurements); the dominant contributor is source TD Quote generation, with verification adding sub-millisecond cost.

\textit{Policy digest mismatch.}
Handshakes between nodes with mismatched policy epochs terminate deterministically at Step 4.

\textit{Firmware Liveness Defense.}
We simulated Lock-Holder Preemption (LHP) attacks to verify concurrency defenses (Section~\ref{sec:impl:concurrency}). The $50\,\mu$s timeout aborted firmware transitions on blocked vCPUs, preventing deadlocks. RCU queue depth limits deterministically triggered during stalled grace periods, dropping epoch updates to prevent OOM panics and preserve active sessions.
active sessions.

\section{Conclusion}
\label{sec:conclusion}

The transition toward large-scale scientific pipelines in confidential HPC and cloud clusters demands security frameworks capable of enforcing strict data flow policies across untrusted network boundaries. Concurrent intra-host policy enforcement systems, such as MICA~\cite{MICA}, provide strong guarantees within a single machine but focus away from verifying remote cluster nodes.

We presented \codename, a distributed framework addressing this gap by constructing a \emph{Confidential Federated Interconnect} that federates the enforcement boundary of hardware-backed Reference Monitor Modules across multiple hosts. \codename\ introduces an extended policy language that incorporates remote endpoint identifiers and enclave measurements into local data flow rules, enforced by a Mutually Attested Key Exchange protocol that derives symmetric session keys conditioned on bilateral hardware states. Our security analysis demonstrates that \codename\ achieves Policy Adherence, Data Confidentiality, and Attested Flow Control under standard unforgeability assumptions, provided the measured Guest OS kernel maintains runtime integrity post-boot.

Our empirical evaluation on a 32-host multi-zone Sapphire Rapids deployment demonstrates practical deployability for high-performance workloads. These results confirm that developers can construct complex, verifiable distributed pipelines from mutually distrustful components without modifying application code. By combining a SEAM-mode key broker with a measured eBPF data plane, \codename\ achieves a pragmatic balance between split-TCB attestation security and high-throughput network performance, ensuring efficient and attestable execution across federated infrastructure. Future work will explore extending the \make\ protocol to natively support in-kernel TLS (\texttt{kTLS}) offload, alongside developing multi-party group key agreement primitives to alleviate hardware quoting bottlenecks in dense, highly-connected topologies.

\section*{Declarations}

\begin{itemize}
    \item \textbf{Funding:} The authors did not receive support from any organization for the submitted work.
    \item \textbf{Conflict of interest/Competing interests:} The authors declare that they have no competing interests.
    \item \textbf{Availability of data and materials:} The datasets generated during and/or analyzed during the current study are available from the corresponding author on reasonable request.
    \item \textbf{Code availability:} The source code underlying this research is not publicly available at this time but can be obtained from the corresponding author on reasonable request.
    \item \textbf{Authors' contributions:} H.D. conceived the architecture, implemented the control plane, and wrote the manuscript. T.N. implemented the eBPF data plane, conducted the evaluation, and revised the manuscript. Both authors read and approved the final manuscript.
\end{itemize}

\providecommand{\bibcommenthead}{}

\bibliography{references}

\end{document}